%% file: main.tex
\documentclass[openany]{nature}
\usepackage[T1]{fontenc}
\usepackage[left]{lineno}
\usepackage{amsthm,amsmath,amssymb}
\usepackage{graphicx}
\usepackage{float}
\usepackage{subfigure}
\usepackage{cases}
\usepackage{mathrsfs}
\usepackage[flushleft]{threeparttable}
\usepackage{pifont}
\usepackage{epstopdf}
\usepackage{xcolor}
\usepackage{hyperref}
\usepackage{mwe,tikz}
\usepackage{bm}
\usepackage{multirow}
\usepackage{longtable}
\usepackage{xspace}
\usepackage{rotating}
\usepackage{CJK}
\usepackage{url}
\usepackage{upgreek}
\usepackage{booktabs}
\usepackage{textcomp}
\usepackage{makecell}
\usepackage{enumitem}
\usepackage{csquotes}
\usepackage{soul}
\usepackage{tabularx}
\usepackage{overpic}
\usepackage{caption}
\usepackage{interval}
\usepackage{float}
\usepackage{threeparttable}
\usepackage{threeparttablex}
\usepackage{subfiles}
\usepackage{array}
\usepackage{makecell}
\usepackage{ulem}

\makeatletter
 
\def\emph#1 {\textit{ #1 } }

\let\saved@includegraphics\includegraphics
\AtBeginDocument{\let\includegraphics\saved@includegraphics}
\renewenvironment*{figure}{\@float{figure}}{\end@float}
\makeatother

\def\@fnsymbol#1{\ensuremath{\ifcase#1\or \dagger\or \ddagger\or
 \mathsection\or \mathparagraph\or \|\or **\or \dagger\dagger
 \or \ddagger\ddagger \else\@ctrerr\fi}}
\makeatother

\setlist[itemize]{leftmargin=*}

\let\oldequation\equation
\let\oldendequation\endequation
\renewenvironment{equation}{\linenomathNonumbers\oldequation}{\oldendequation\endlinenomath}

\let\oldalign\align
\let\oldendalign\endalign
\renewenvironment{align}{\linenomathNonumbers\oldalign}{\oldendalign\endlinenomath}

\let\oldgather\gather
\let\oldendgather\endgather
\renewenvironment{gather}{\linenomathNonumbers\oldgather}{\oldendgather\endlinenomath}

\newcommand{\araa}{Annu Rev Astron Astrophys}

\newcommand{\ana}{Astron Astrophys}

\newcommand{\aj}{Astron J}

\newcommand{\apj}{Astrophys J}
\newcommand{\apjl}{Astrophys J Lett}
\newcommand{\apjs}{Astrophys J Suppl Ser}

\newcommand{\aap}{Astron Astrophys}

\newcommand{\jcap}{J Cosmol Astropart Phys}

\newcommand{\mnras}{Mon Not R Astron Soc}
\newcommand{\nat}{Nature}

\newcommand{\physrep}{Phys Rep}

\newcommand{\prd}{Phys Rev D}

\newcommand{\pasp}{Publ Astron Soc Pac}

\newcommand{\ssr}{Space Sci Rev}

\newcommand{\src}{EP250702a}
\newcommand{\sw}{Sw J1644+57}
\title{A fast powerful X-ray transient from possible tidal disruption of a white dwarf}

\author{Dongyue Li$^1$\thanks{These authors contributed equally to this work\\Corresponding authors:}, Wenda Zhang$^{1*}$\thanks{wdzhang@nao.cas.cn}, Jun Yang$^{2*}$, Jin-Hong Chen$^{3,4,5*}$, Weimin Yuan$^{1,6}$\thanks{wmy@nao.cas.cn}, Huaqing Cheng$^1$, Fan Xu$^7$, Xinwen Shu$^7$, Rong-Feng Shen$^{8,9}$, Ning Jiang$^{10}$, Jiazheng Zhu$^{10}$, Chang Zhou$^{11}$, Weihua Lei$^{11}$, Hui Sun$^1$, Chichuan Jin$^{1,6,12}$\thanks{ccjin@nao.cas.cn}, Lixin Dai$^{3,4}$\thanks{lixindai@hku.hk}, Bing Zhang$^{3,4}$, Yu-Han Yang$^{13}$, Wenjie Zhang$^{1}$, Hua Feng$^{14}$, Bifang Liu$^{1,6}$, Hongyan Zhou$^{10,15}$, Haiwu Pan$^1$, Mingjun Liu$^1$, Stéphane Corbel$^{16}$, Sitha K. Jagan$^{17}$, Maria Cristina Baglio$^{18}$, Christopher R. Burns$^{19}$, Floriane Cangemi$^{20}$, Chun Chen$^{8,9,21}$, Yehao Cheng$^{22}$, Alexis Coleiro$^{20}$, Francesco Coti Zelati$^{23,24}$, Sourya R. Das$^{16}$, Zhongnan Dong$^{8,20,1}$, Luis Galbany$^{23,24}$, Noa Grollimund$^{16}$, Daniel Kelson$^{19}$, Dong Lai$^{25,26}$, Xia Li$^{8,9}$, Yuan Liu$^{1}$, Alessio Marino$^{23,24}$, Brenna Mockler$^{19}$, Paul O'Brien$^{27}$, Erlin Qiao$^{1,6}$, Nanda Rea$^{23,24}$, Resmi$^{17}$, Jérome Rodriguez$^{16}$, Richard Saxton$^{28}$, Luming Sun$^{7}$, Lian Tao$^{13}$, Tinggui Wang$^{10}$, Yilong Wang$^{1,23,24}$, Xuefeng Wu$^{29}$, Dong Xu$^1$, Yijia Zhang$^{30}$, Guoying Zhao$^{8}$, Congying Bao$^{1}$, Zhiming Cai$^{31}$, Yehai Chen$^{31}$, Yong Chen$^{14}$, Bertrand Cordier$^{32}$, Chenzhou Cui$^{1}$, Weiwei Cui$^{14}$, Zhou Fan$^{1,6}$, He Gao$^{12,33}$, Giancarlo Ghirlanda$^{18,34}$, Ju Guan$^{14}$, Dawei Han$^{14}$, Jinxin Hao$^{1}$, Jingwei Hu$^{1}$, Maohai Huang$^{1}$, Yong-Feng Huang$^{35,36}$, Shumei Jia$^{14}$, Ge Jin$^{37}$, Stefanie Komossa$^{38}$, Chengkui Li$^{14}$, Zhixing Ling$^{1,6,12}$, Congzhan Liu$^{14}$, Heyang Liu$^{1}$, Huaqiu Liu$^{31}$, Fangjun Lu$^{14}$, Kirpal Nandra$^{39}$, Jan-Uwe Ness$^{40}$, Arne Rau$^{39}$, Jeremy Sanders$^{39}$, Liming Song$^{14}$, Roberto Soria$^{41,42}$, Shengli Sun$^{43}$, Xiaojin Sun$^{43}$, Yunyin Tan$^{44}$, Eleonora Troja$^{13}$, Sixiang Wen$^{1}$, Haitao Xu$^{44}$, Changbin Xue$^{44}$, Yongquan Xue$^{10}$, Yi-Han Iris Yin$^{3,4}$, Chen Zhang$^{1}$\thanks{chzhang@bao.ac.cn}, Shuang-Nan Zhang$^{14}$, Yonghe Zhang$^{31}$}

\begin{document}
\maketitle

\begin{affiliations}
\item{National Astronomical Observatories, Chinese Academy of Sciences, Beijing 100101, China}
\item{Institute for Astrophysics, School of Physics, Zhengzhou University, Zhengzhou 450001, China}
\item{The Hong Kong Institute for Astronomy and Astrophysics, The University of Hong Kong, Hong Kong 33983, China}
\item{Department of Physics, The University of Hong Kong, Hong Kong 999077, China}
\item{Shenzhen Institute of Research and Innovation, The University of Hong Kong, Shenzhen 518057, China}
\item{School of Astronomy and Space Science, University of Chinese Academy of Sciences, Beijing 100049, China}
\item{Department of Physics, Anhui Normal University, Wuhu 241002, China}
\item{School of Physics and Astronomy, Sun Yat-sen University, Zhuhai 510982, China}
\item{CSST Science Center for the Guangdong-Hong Kong-Macau Greater Bay Area, Zhuhai 519082, China}
\item{Department of Astronomy, University of Science and Technology of China, Hefei 230026, China}
\item{Department of Astronomy, School of Physics, Huazhong University of Science and Technology, Wuhan 430074, China}
\item{Institute for Frontier in Astronomy and Astrophysics, Beijing Normal University, Beijing 102206, China}
\item{Department of Physics, University of Rome ``Tor Vergata'', via della Ricerca Scientifica 1, Rome I-00133, Italy}
\item{Key Laboratory of Particle Astrophysics, Institute of High Energy Physics, Chinese Academy of Sciences, Beijing 100049, China}
\item{Polar Research Institute of China, Shanghai 200136, China}
\item{Université Paris Cité, Université Paris-Saclay, CEA, CNRS, AIM, Gif-sur-Yvette F-91191, France}
\item{Indian Institute of Space Science and Technology, Trivandrum 695547, India}
\item{INAF--Osservatorio Astronomico di Brera, Via Bianchi 46, I-23807 Merate (LC) I-23807, Italy}
\item{The Observatories of the Carnegie Institution for Science, Pasadena, CA 91101, USA}
\item{Université Paris Cité, CNRS, Astroparticule et Cosmologie, Paris F-75013, France}
\item{Dipartimento di Fisica, Universit$\grave{a}$ di Napoli “Federico II”, Compl. Univ. di Monte S. Angelo, Via Cinthia I-80126, Italy}
\item{South-Western Institute for Astronomy Research, Yunnan University, Kunming 650504, China}
\item{Institute of Space Sciences (ICE, CSIC), Campus UAB, Carrer de Can Magrans s/n, Barcelona E-08193, Spain}
\item{Institut d’Estudis Espacials de Catalunya (IEEC), Barcelona E-08034, Spain}
\item{Tsung-Dao Lee Institute, Shanghai Jiao-Tong University, Shanghai 201210, China}
\item{Center for Astrophysics and Planetary Science, Department of Astronomy, Cornell University, Ithaca, NY 14853, USA}
\item{School of Physics and Astronomy, University of Leicester, Leicester LE1 7RH, UK}
\item{Telespazio UK for ESA, ESAC, Apartado 78, Villanueva de la Ca\~{n}ada 28691,  Spain}
\item{Purple Mountain Observatory, Chinese Academy of Sciences, Nanjing 210023, China.}
\item{Department of Astronomy, Tsinghua University, Beijing 100084, China}
\item{Key Laboratory for Satellite Digitalization Technology, Innovation Academy for Microsatellite, Chinese Academy of Sciences, Shanghai 201304, China}
\item{CEA Paris-Saclay, Irfu/D'epartement d’Astrophysique, Gif sur Yvette 9111, France}
\item{School of Physics and Astronomy, Beijing Normal University, Beijing 100875, China}
\item{INFN – Sezione di Milano-Bicocca, Piazza della Scienza 3, Milano (MI) I-20146, Italy}
\item{School of Astronomy and Space Science, Nanjing University, Nanjing 210023, China}
\item{Key Laboratory of Modern Astronomy and Astrophysics (Nanjing University), Ministry of Education, Nanjing 210023, China}
\item{North Night Vision Technology Co., LTD, Nanjing 210008, China}
\item{Max-Planck-Institut fuer Radioastronomie, Auf dem Huegel 69, Bonn 53121, Germany}
\item{Max-Planck-Institut f\"ur extraterrestrische Physik, Giessenbachstrasse 1, Garching 85748, Germany}
\item{European Space Agency, European Space Astronomy Centre, Madrid E-28692, Spain}
\item{INAF-Osservatorio Astrofisico di Torino, Pino Torinese I-10025, Italy}
\item{Sydney Institute for Astronomy, School of Physics A28, The University of Sydney, NSW 2006, Australia}
\item{The Shanghai Institute of Technical Physics of the Chinese Academy of Sciences, Shanghai 200083, China}
\item{National Space Science Center, Chinese Academy of Sciences, Beijing 100190, China}
\end{affiliations}

\begin{abstract}
{\bf Abstract\\}
Stars getting close enough to black holes (BHs) can be torn apart by strong tidal forces, producing electromagnetic flares. To date, more than 100 tidal disruption events (TDEs) have been observed, each involving invariably normal gaseous stars whose debris falls onto the BH, sustaining the flares over years. White dwarfs (WDs), which are the most prevalent compact stars and a million times denser--and therefore tougher--than gaseous stars, can only be disrupted by intermediate-mass black holes (IMBHs) of $10^{2}$--$10^{5}$ solar masses. WD-TDEs are considered to generate more powerful and short-lived flares, but their evidence has been lacking. Here we report observations of a fast and luminous X-ray transient \src{} detected by Einstein Probe. Its one-day-long X-ray peak as luminous as $10^{47-49}$\,erg\,s$^{-1}$ showed strong recurrent flares with hard spectra extending to several tens of MeV gamma-rays, as detected by Fermi/GBM and Konus-Wind, indicating relativistic jet emission. The jet's X-rays dropped sharply from $3 \times 10^{49}$\,erg\,s$^{-1}$ to around $10^{44}$\,erg\,s$^{-1}$ within 20 days (10 days in the source rest frame). These characteristics are inconsistent with any previously known transient phenomena. We suggest that this fast-evolving event over the unprecedentedly short timescale arises likely from disruption of a WD by an IMBH. At late times, a soft component progressively dominates the X-ray spectrum, reaching a luminosity as high as 10$^{44}$ erg/s, which is consistent with being extreme super-Eddington emission from an accretion disk expected to form in an IMBH-WD TDE. WD-TDEs open a new window for investigating the elusive IMBHs and their surrounding stellar environments, and they are prime sources of gravitational waves in the band of space-based interferometers.

\end{abstract}

\bigskip
\noindent{\bf keywords:} X-ray transient; Intermediate-mass black hole; Tidal disruption event; White dwarf; Tianguan Einstein Probe

\section{Introduction}
Intermediate-mass black holes (IMBHs), with masses in the range of $10^2$--$10^5~M_\odot$, occupying the mass gap between stellar-mass (a few to tens of solar masses) and supermassive black holes (SMBHs, with masses above $10^5~M_\odot$), represent a critical missing link in the cosmic evolution of SMBHs\cite{Volonteri2012, Greene2020}. Detections of IMBHs offer us great insights into the seeding and growth of SMBHs\cite{inayoshi_assembly_2020}. Tremendous efforts have been put on hunting IMBHs, especially in dwarf galaxies, globular clusters, and active nuclei of low-mass galaxies. However, very few candidates have been detected so far (e.g., Omega Centauri\cite{Haberle2024}).

When a star passes close enough to a black hole, it can be tidally disrupted and accreted, resulting in an electromagnetic outburst\cite{Hills1975, Rees1988, Komossa2015}, which is known as tidal disruption events (TDEs). TDEs provide a unique probe for quiescent black holes which are otherwise difficult to detect, as well as an ideal laboratory for studying accretion around them. More than 100 TDEs have been discovered as X-ray, optical, and UV transients on timescales of months to years\cite{Saxton2021,Gezari2021}. Tidal disruption events by IMBHs offer a novel approach to detect new IMBHs, to study their formation and evolution, and to investigate black hole accretion physics in this elusive mass regime\cite{Greene2020}.

It is particularly intriguing when the disrupted star is a white dwarf (WD) rather than a main sequence star, which is typically the focus of observation. WDs are the final evolutionary stage of most low- and intermediate-mass stars ($\lesssim8~M_{\odot}$). Compared with main sequence stars, WDs possess denser cores and stronger magnetic fields. Due to the compactness of WDs, they can only be disrupted by black holes with masses $\lesssim10^5~M_{\odot}$, making detections of WD-TDEs a smoking-gun evidence of IMBHs.
It has been predicted that WD-IMBH TDEs will give rise to short-lived bright flares, potentially launch a relativistic jet, and may even be accompanied by a thermonuclear explosion triggered by the tidal compression of the WD\cite{Maguire2020, MacLeod2016}. 
However, no clear evidence for WD-IMBH TDE has been found hitherto, though this scenario was invoked to explain some of the observed properties in a few transients previously detected with short timescales (e.g., Refs.\cite{Clausen2011, Jonker2013, Kuin2019}).

In this paper, we report the discovery of a fast and luminous X-ray transient \src{} by the Einstein Probe (EP)\cite{Yuan2022} mission and follow-up observations and investigations. Its unique observed properties, including a long-duration X-ray and gamma-ray outburst, fast evolution of the transient X-ray light curve, an extremely high isotropic X-ray luminosity and rapid variability, the emergence of a soft component at a late stage of the X-ray spectral evolution, do not resemble those of any transients known previously. 
Instead, they all suggest the transient \src{} to most likely arise from tidal disruption of a WD by an IMBH producing a relativistic jet. 

Launched on 9 January 2024, EP\footnote{The Einstein Probe is a space mission led by the Chinese Academy of Sciences (CAS), in collaboration with the European Space Agency (ESA), the Max Planck Institute for Extraterrestrial Physics (MPE) in Germany, and the French Space Agency (CNES).} is an interdisciplinary X-ray observatory with the main science goals being to detect cosmic X-ray transients. It carries 12 identical Wide-field X-ray Telescope (WXT) modules which utilize novel lobster-eye micro-pore optics (MPO) for X-ray focusing imaging. They achieve both a large instantaneous field of view (FoV; $\sim$3,850 square degrees) and relatively high sensitivity ($(2$--$3)\times 10^{-11}$\,erg\,s$^{-1}$\,cm$^{-2}$\ for 1~ks exposure), as well as good spatial resolution ($\sim$5$^\prime$, full width at half maximum) across the entire FoV.  
These unique features make EP-WXT a powerful instrument for discovering X-ray transients and monitoring known sources. 
WXT's wide-field monitoring capability is complemented by an onboard Follow-up X-ray Telescope (FXT), with a FoV of 1\,degree, for quick and deep follow-up observations.

\section{Observations and Data Analysis}
\begin{figure}
\centering
\includegraphics[width=\textwidth]{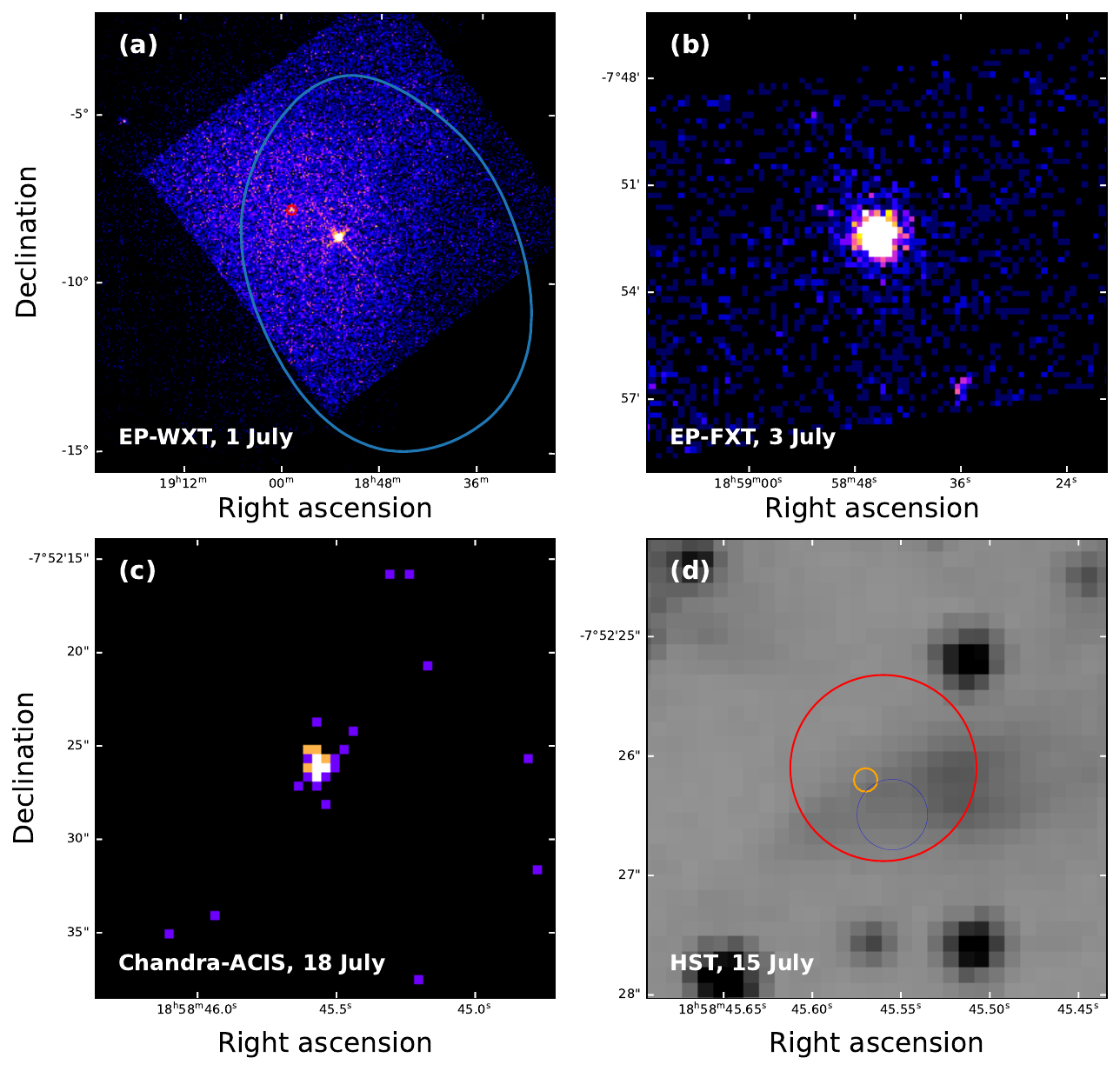}
\caption{The field of EP250702a in EP, Chandra, and HST imaging. (a)--(c), X-ray images of EP250702a taken by EP-WXT, EP-FXT, and {\it Chandra}, respectively. The position of EP250702a is indicated with red circles, and the blue contour shown in (a) gives the 1$\sigma$ position uncertainty of GRB 250702B. (d), Optical image of EP250702a observed by HST. The red, orange, and blue ellipses indicate the positions of the X-ray, near-infrared (NIR), and radio counterparts of EP250702a, respectively. The NIR position and its uncertainties are obtained from VLT observations\cite{Levan2025}. The radio position and its uncertainties are obtained from MeerKAT observations (Supplementary material, observations and data reduction).}
\label{fig:opt_img}
\end{figure}

On 2 July 2025, the EP-WXT detected an X-ray transient exhibiting strong flaring activity, designated EP250702a\cite{GCN40906}. 
The X-ray source was well localized with an uncertainty of only 2.4$^\prime$ (90\% C.L. statistical and systematic), 
thanks to the good spatial resolution of WXT. 
The source is spatially and temporally consistent with multiple gamma-ray flares, designated GRB 250702D, B, and E (hereafter GRB 250702B), each lasting several hundred seconds and together spanning a period of more than three hours, as detected by the Gamma-ray Burst Monitor (GBM) onboard the Fermi Gamma-ray Space Telescope, with spatial uncertainties of $7.8^\circ$--$14.7^\circ$\cite{GCN40891, GCN40931}. 
Associated hard X-ray and gamma-ray flaring activity from this source was also detected by the Monitor of All-sky X-ray Image (MAXI)\cite{GCN40910}, Konus-Wind\cite{GCN40914}, and the Space Variable Objects Monitor (SVOM)\cite{GCN40923}. A backward search in the EP-WXT data revealed the emergence of its X-ray emission already since 01:40:26 Coordinated Universal Time (UTC) on 1 July 2025 (defined as the trigger time), which is about one day before the onset of the reported gamma-ray flares (Fig.~\ref{fig:opt_img}a; Supplementary material, observations and data reduction). A targeted backward search in the Fermi/GBM data also uncovered an untriggered weak flare at 11:55:18 UTC on 1 July 2025, whose localization is broadly consistent with that of EP250702a (Fig.~\ref{fig:gbm_balrog} online; Supplementary material, observations and data reduction). 

The detections of \src{} and the associated GRB 250702B have triggered extensive campaigns of multi-wavelength follow-up observations. 
Based on the promptly well determined WXT position, EP-FXT observed \src{} in a quick follow-up on 3 July (approximately two days after the trigger) and further pinned down the source with an uncertainty of 10$^{\prime\prime}$\cite{GCN40917} (Fig.~\ref{fig:opt_img}b), thereby facilitating subsequent multi-wavelength follow-ups. The most precise X-ray localization was provided by a 15-ks observation with the {\it Chandra} Advanced CCD Imaging Spectrometer (ACIS) on 18 July (PI: D.-Y. Li), as shown in Fig.~\ref{fig:opt_img}c. The coordinates of the X-ray source are right ascension (R.A.) = $18{\rm h}58{\rm m}45.56{\rm s}$ and declination (Dec.) = $-7^{\circ}52^{\prime}26.1^{\prime\prime}$ (J2000), with a $1\sigma$ uncertainty of $0.79^{\prime\prime}$ that includes both statistical and systematic errors. A decaying near-infrared (NIR) counterpart, with an extremely red color and located off the nucleus of an underlying galaxy, was identified in Very Large Telescope (VLT) observations starting from 3 July, and a radio counterpart was detected with MeerKAT observations starting on 4 July\cite{Levan2025}. These positions are both consistent with the X-ray localization (Fig.~\ref{fig:opt_img}d), pinning down the transient at the outskirt of an external galaxy. The redshift of EP250702a was determined to be $z = 1.036 \pm 0.004$ based on a JWST spectroscopic observation\cite{gompertz2025jwst}. We performed high-cadence monitoring observations of \src{} by EP-FXT until about 40 days after the trigger, when the source became no longer detectable.

\begin{figure}
\centering
\includegraphics[width=0.9\textwidth]{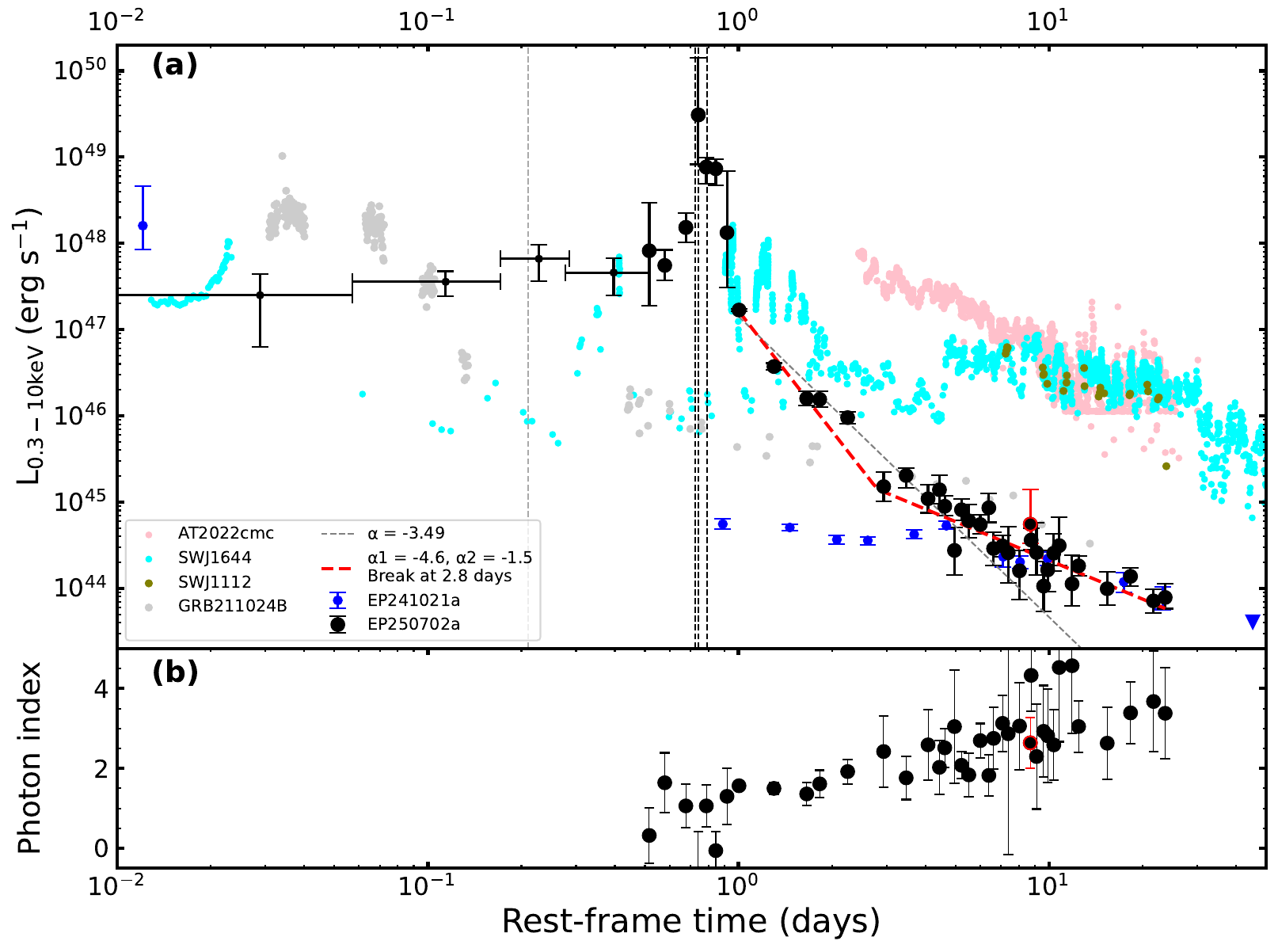}
\caption{{Long-term X-ray light curve and spectral evolution of EP250702a.} (a), Comparison of EP250702a with other X-ray transients, including jetted TDEs (\sw{}, Swift J1112.2-8238, AT2022cmc), an ultra-long GRB (GRB 211024B) and a jetted TDE candidate EP241021a\cite{Shu2025}, on their X-ray light curves. The trigger time of EP250702a is set to be the start time of the first WXT observation when EP250702a begins to emerge (Supplementary material, observations and data reduction), about one day before the first Fermi/GBM trigger. The data points of EP250702a with red borders and red error bars represent measurement from the {\it Chandra} observation. The overlaid gray dashed line represents the single powerlaw decay fit, and the overlaid red dashed curve represents the two-segment power-law decay fit to the long-term X-ray light curve of EP250702a. The vertical dashed black lines denote the trigger times of the Fermi/GBM flares, and the vertical dashed gray one gives the time of the untriggered weak Fermi/GBM flare (Supplementary material, observations and data reduction). 
(b), The photon indices for EP250702a derived from absorbed power-law spectral fitting to the X-ray spectra. All error bars represent 1$\sigma$ uncertainties.}
\label{fig:xray_evol}
\end{figure}

By combining the WXT and FXT data, we obtain a densely sampled, complete X-ray light curve of \src{} (Fig.~\ref{fig:xray_evol}a). 
The light curve shows an initial outburst phase lasting for $\sim1$ day and followed by a rapid decay (the ``afterglow'' phase) with the flux declining by more than five orders of magnitude within 20 days (about 10 days in the source rest frame). 
Assuming isotropic emission, the outburst reached an apparent X-ray luminosity as high as several times $10^{47}$\,erg\,s$^{-1}$ at the first detection of \src{} on 1 July.
This hard non-thermal emission (see below) persisted relatively steadily for about 10 hours, followed by several hours (a factor of $(1+z)\sim 2$ shorter in the source rest frame) of intense X-ray flaring activity with variability amplitudes exceeding one order of magnitude and reaching $3\times 10^{49}$\,erg\,s$^{-1}$ at peaks. Concurrently, three exceptionally bright and high-energy gamma-ray flares were detected. 
Prior to the start of the first-detection observation, the source position was not covered by WXT for 5 hours. Stacking the data acquired earlier yields an upper limit of $3.5\times 10^{46}$\,erg\,s$^{-1}$. 
Thus the initial rising phase from the onset of the event to the outburst state  cannot be constrained.

\begin{figure}
\centering
\begin{tabular}{cc}
\begin{overpic}[width=0.56\textwidth]{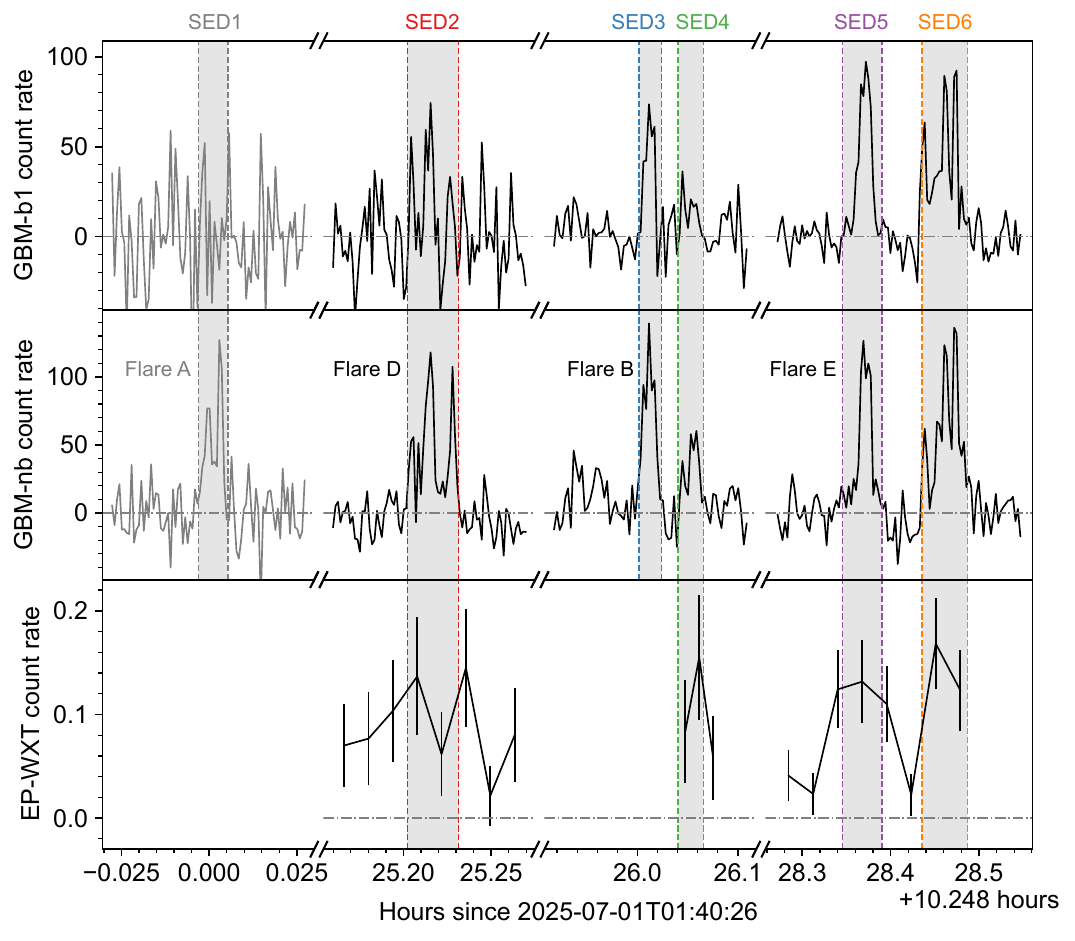}\put(0, 87){\bf (a)}\end{overpic} &
\begin{overpic}[width=0.38\textwidth]{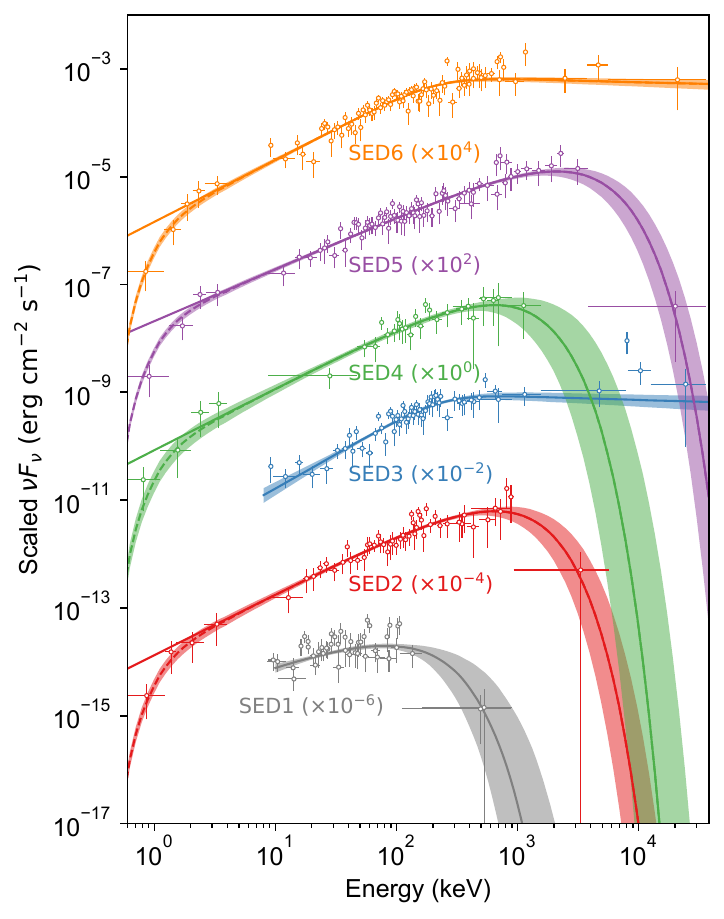}\put(0, 97){\bf (b)}\end{overpic} \\
\end{tabular}
\caption{Light curves and spectral energy distributions (SEDs) during the flaring episodes of EP250702a. (a), Light curves observed during the flaring episodes by the BGO detector of Fermi/GBM (top panel), the NaI detector of Fermi/GBM (middle panel), and EP-WXT (bottom panel). Six distinct flaring episodes are indicated by grey shaded regions with color-coded boundaries. (b), SEDs corresponding to the six flaring episodes. The SEDs are derived from the joint spectral fits over the time intervals indicated by the labels. Solid and dashed lines represent the best-fit unabsorbed and absorbed models, respectively. Error bars on the data points denote the 1$\sigma$ confidence level, and the shaded regions around the best-fit lines indicate the corresponding 1$\sigma$ confidence bands.}
\label{fig:xray_gamma}
\end{figure}

The extremely high apparent isotropic luminosity of $10^{47-49}$\,erg\,s$^{-1}$, the hard non-thermal spectrum from X-ray to gamma-ray, and the rapid variability of the flares (see below) indicate that the X-ray emission is highly beamed, originating from a transient relativistic jet. The emergence of the X-rays about one day before the gamma-ray flares, together with the exceptionally long gamma-ray flare durations
(spanning about 2 hours and possibly extending beyond 14 hours, in the source rest frame, if the untriggered gamma-ray flare on 1 July was also associated with \src{}), argues against gamma-ray burst (GRB) scenarios, even those involving ultra-long GRBs (Figs.~\ref{fig:xray_evol}a and \ref{fig:xray_gamma}a). The behavior of \src{} neither fits any known Galactic transients. Instead, the extremely prolonged outburst with highly variable fluctuations/flares closely resembles the rare class of \textit{jetted} TDEs, such as SW J1644 and AT2022cmc (over-plotted on Fig.~\ref{fig:xray_evol}a), except that they have even longer flare timescales. In such events, accretion of fallback stellar debris powers relativistic outflows, producing luminous and energetic non-thermal outbursts in the X-ray and gamma-ray bands. 

The short variability timescale of the early outburst phase indicates an IMBH powering this event. The light curves during the flaring episodes observed by Fermi/GBM reveal a short minimum variability timescale of 1.5\,s, corresponding to 0.74\,s in the source rest frame (Table~\ref{tab:x_gamma_prop} online; Supplementary material, timing analysis). Such a brief timescale places an upper limit on the characteristic size of the source emitting region, $R_{\rm E} < 2.2 \times 10^{10}~{\rm cm}$, which implies an IMBH with a mass of $M_{\rm BH} < 7.5 \times 10^4~{M_\odot}$ given that $R_{\rm E}$ cannot be smaller than the radius of the black hole event horizon for a Schwarzschild black hole. This is further supported by its off-nuclear position within its host galaxy. We thus suggest that \src{} provides likely the clearest evidence for a jetted TDE produced by an IMBH.

The X-ray flux of this event began to decay much earlier, with a significantly faster decaying rate than all previously documented jetted TDEs (Fig.~\ref{fig:xray_evol}a). Furthermore, its decay timescale is also substantially shorter than those of the three previously reported IMBH-TDE candidates, which typically exhibit years-long decay profiles\cite{Lin2018, Jin2025, Yao2025, He2021}. We also note that its X-ray luminosity at the peaks of the outburst is at least 1--2 orders of magnitude higher than those of all the known jetted TDEs. The short evolutionary timescale, together with the highest peak luminosity in X-ray, points to a WD, rather than a main-sequence star, as the disrupted star, since the fallback timescale for the latter would be much longer (tens of days or even longer)\cite{Chen2018}. Our first-order theoretical scaling analysis further confirms that the rapid decay timescale of \src{} is consistent with the tidal disruption of a low-mass WD by an IMBH (Figs.~\ref{fig:IMBH_WD_modelling}--\ref{fig:rTDE_WD_modelling} online; Supplementary material, theoretical modelling). 

Compared with other jetted TDEs, \src{} also stands out by producing high-energy flares extending from the X-ray to the gamma-ray regime of several tens of ${\rm MeV}$.  (Fig.~\ref{fig:xray_gamma}a). We extracted the 0.5\,keV--38\,MeV broad-band spectra of \src{} from the EP-WXT and Fermi/GBM data for the flaring episodes, and found that the spectra are best fit with a non-thermal spectral model, characterized by a low-energy photon spectral index of approximately $-1$, a peak energy reaching up to $2~{\rm MeV}$, and a high-energy power-law component extending to several tens of ${\rm MeV}$ (Fig.~\ref{fig:xray_gamma}b; Supplementary material, spectral analysis). The total fluence during the flaring phase, inferred from spectral fitting, is $\gtrsim (1.7\pm0.1)\times10^{-4}~{\rm erg~cm^{-2}}$, corresponding to an isotropic-equivalent energy of $\gtrsim (5.0\pm0.3)\times10^{53}~{\rm erg}$ at a redshift of $z=1.036$. This unprecedented high-energy peak at the gamma-ray band significantly exceeds the soft and hard X-ray regime typically observed in known jetted TDEs\cite{Burrows2011}. Based on the absence of a cutoff feature in the broad-band spectra, above which the pair-production opacity may reach unity, together with the observed variability timescale, we are able to place a lower limit on the jet bulk Lorentz factor under the assumption of the internal shock model. This yields $\Gamma_{\rm j} \gtrsim 56$ (Supplementary material, constraints on the jet bulk Lorentz factor), which is similar to the values reported for previous jetted TDEs\cite{Burrows2011, Bloom2011, Andreoni2022}. 

The WD-IMBH TDE scenario offers a natural explanation for the gamma-ray emission of \src{} that is unique among jetted TDEs. The radiation mechanisms responsible for the flaring X-ray and gamma-ray emission of jetted TDEs remain unclear. Possible models include synchrotron radiation\cite{Burrows2011, Yao2024}, synchrotron self-Compton (SSC\cite{Bloom2011, Pasham2023}), and external inverse-Compton\cite{Bloom2011}. Although a larger Lorentz factor could lead to a higher peak energy, this is not the case for \src{} as its Lorentz factor is comparable to other jetted TDEs. Another factor that determines the peak energy is the strength of the magnetic field if the radiation mechanism responsible for the high-energy emission is synchrotron or SSC. In this context, a magnetized WD as the disrupted star for \src{} can offer a stronger magnetic field for the accretion disk that forms after the tidal disruption, thus pushing the peak energies to the gamma-ray band.

The extremely high X-ray luminosity at the first WXT detection of \src{} indicates that the jet had already been launched by then. 
The initial flaring activity phase, with strong variabilities reaching peak luminosities two orders of magnitudes higher, lasted for almost one day. This implies that the process of launching and intensifying the jet in \src{} takes longer compared with the typical debris fallback time scale expected for a WD-TDE by an IMBH. 
This may indicate that the disk formation and the magnetic flux accumulation processes\cite{Tchekhovskoy2014}, both needed for launching relativistic jets through the Blandford-Znajek process\cite{Blandford1977}, likely happen over timescales much longer than the debris fallback timescale.

\begin{figure}
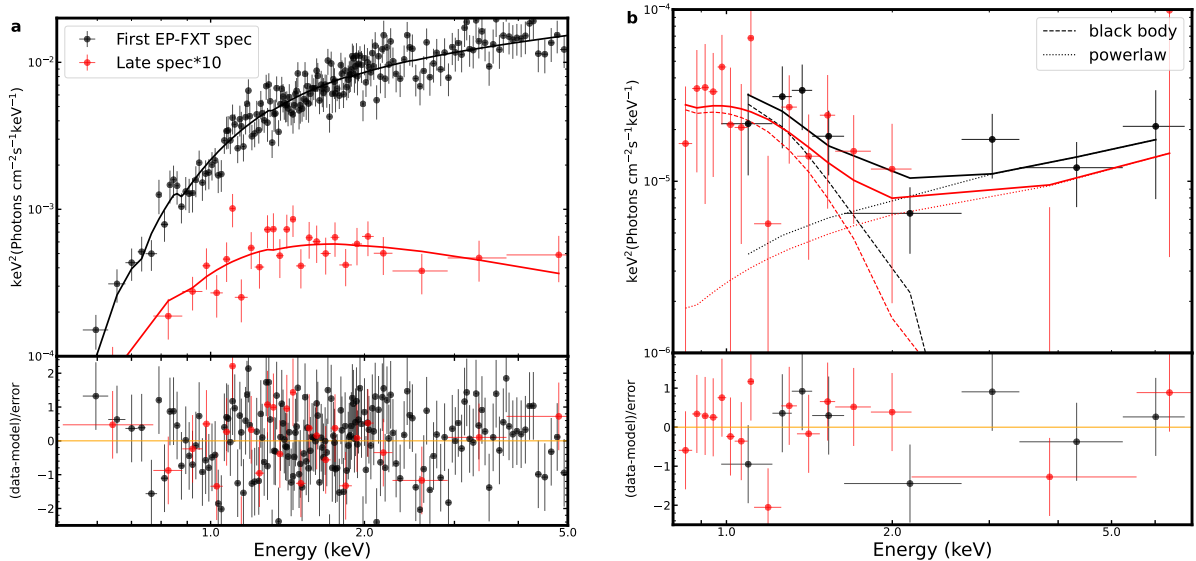

\centering
\begin{tabular}{cc}
\begin{overpic}[width=0.47\textwidth]{figures/main_fig4a.pdf}\put(-1, 92){\bf (a)}\end{overpic} &
\begin{overpic}[width=0.47\textwidth]{figures/main_fig4b.pdf}\put(-3, 92){\bf (b)}\end{overpic} \\
\end{tabular}
\caption{X-ray spectra of EP250702a detected by EP and Chandra. (a), Evolution of X-ray spectra for EP250702a showing significant spectral softening trend in later observations. The spectrum taken from the first FXT observation on 2 July is shown in black points, and the late-time stacked spectrum (with the data taken from 12 July to 15 July, when the averaged spectral photon index is greater than 2) is shown in red. The late-time stacked spectrum has been scaled by a factor of 10 for visual clarity. Both spectra were fitted with an absorbed power-law model, and the fitting residuals were shown in the lower panel. (b), Joint X-ray spectral fitting of simultaneous {\it Chandra} and FXT observations, taken on 18 July, with the best-fit absorbed blackbody (dashed line) and power-law (dotted line) model components, as well as their combined model \textit{const*tbabs*ztbabs*(bbody+powerlaw)} (solid line) (Supplementary material, spectral analysis). The fitting residuals are shown in the lower panel. The black and red points represent the {\it Chandra} and FXT observations, respectively. All error bars represent 1$\sigma$ uncertainties.}
\label{fig:xray_lc_spec}
\end{figure}

The flux decay is accompanied by dramatic spectral evolution, beginning with an extremely hard early-phase spectrum that progressively hardens during the period of flaring activity and then softens significantly during late-phase decay. This hard-to-soft transition is obvious by looking at the change of the photon index $\Gamma$ of the spectrum that is measured by fitting an absorbed power-law model to the soft X-ray energy spectrum, which shows a clear transition from $\Gamma \approx 1$ to $\Gamma \approx 3$ (Fig.~\ref{fig:xray_evol}b), and evident by comparing X-ray spectra of early and late phases (Fig.~\ref{fig:xray_lc_spec}a). Interestingly, the time from when the spectrum started to soften (several days after trigger) roughly coincides with a transition from steep ($f_{\rm X}\propto t^{-4.6}$) to shallow ($f_{\rm X} \propto t^{-1.5}$) decay in the X-ray flux evolution (Fig.~\ref{fig:xray_evol}a).

The spectral softening is accompanied by and probably is due to the emergence and gradual dominance of a thermal component during the late stage of the event. The break in the decay profile therefore indicates that the thermal component decays slower than the jet. The thermal component is significantly detected in the joint fit to the simultaneous {\it Chandra} and FXT observations taken about 17\,days after trigger (Fig.~\ref{fig:xray_lc_spec}b; Supplementary material, spectral analysis).  
The composite model combining blackbody that is used to describe the thermal component and the power-law, as demonstrated in Fig.~\ref{fig:xray_lc_spec}b, yields a blackbody temperature of $kT = 169^{+37}_{-37}~\mathrm{eV}$ (typical of TDEs\cite{Komossa2002}), and a power-law index $\Gamma = 1.4^{+0.8}_{-0.8}$ that is consistent with the early-phase value. The unabsorbed flux of the thermal component is estimated to be 1.7 $\times 10^{-13}$\,erg\,s$^{-1}$\,cm$^{-2}$, and the corresponding luminosity is 9.8 $\times 10^{44}$\,erg\,s$^{-1}$.

The observed luminosity of the thermal component corresponds to $\sim$100\,$L_{\rm Edd}$ (the Eddington limit of a pure-hydrogen accretion disk) of a black hole of $7.5\times 10^4~M_\odot$ in late time. While simulations of super-Eddington disks typically exhibit luminosities of $\lesssim 10~L_{\rm Edd}$\cite{sadowski_numerical_2014, jiang_super-eddington_2019}, WD-IMBH TDEs likely produce accretion disks with very high Eddington ratio and magnetization, which can lead to much higher disk luminosities\cite{McKinney2015}.
Furthermore, this tension can be alleviated for a WD because, in the case of a fully ionized carbon-oxygen WD debris, the Eddington luminosity $L_{\rm Edd,CO}$ is twice $L_{\rm Edd}$ due to its lower electron scattering opacity. In the context of WD-IMBH TDEs, thermonuclear outbursts can occur, especially during deep tidal penetration events. As highlighted by MacLeod et al. (2016)\cite{MacLeod2016}, such outbursts are suggested to subtly affect the binding energy distribution of the debris, however their impact remains secondary to the gravitational dynamics that dominate accretion. Additionally, the outbursts themselves generate optical emission with peak luminosities around 10$^{42}~\rm erg~s^{-1}$, which is significantly lower than the observed X-ray luminosity of 10$^{44}~\rm erg~s^{-1}$. The total luminosity from these thermonuclear events is estimated to be $\sim 10^{41}~(M_{\rm BH}/10^3 M_\odot)~\rm erg~s^{-1}$ \cite{rosswog_tidal_2009}, with only a small fraction emerging in the soft X-ray band. Given that this luminosity falls below the detection limit at the redshift of EP250702a, the thermonuclear component is unlikely to significantly influence the observed accretion-driven light curve.

\section{Discussion and Conclusion}
EP250702a exhibits rapid variability during early phases (hundred-second timescales), while the X-ray spectra photon index evolution demonstrates hardening-softening behavior unparalleled in comparison samples. This distinct timing signature, soft X-ray emission beginning about one day before the first Fermi/GBM trigger, provides an exceptional case study for IMBH accretion dynamics. EP250702a represents the first jetted TDE exhibiting late-phase spectral softening, characterized by unprecedented rapid evolution (more than five orders of magnitudes flux decline in 10 days in the source rest frame) and emergent thermal components consistent with an IMBH tidally disrupting a WD while launching relativistic jets. By modelling the near-infrared and radio light curves within the first month of its evolution in the context of synchrotron afterglow emission from the interaction of jet with the ambient medium, we find an isotropic kinetic jet energy of $E_{\rm K, iso} \approx 4.1\times10^{53}$ erg (Fig.~\ref{fig:AG} online; Supplementary material, afterglow modelling), which is comparable to that of the previously reported jetted TDEs\cite{Zhou2024}.

We have examined several alternative models to interpret the event EP250702a (Supplementary material, other possible scenarios), including micro-tidal disruption events ($\mu$TDEs) involving stellar-mass black holes\cite{Beniamini2025} and scenarios wherein a stellar-mass black hole spirals into a helium-star envelope via unstable mass transfer or common-envelope evolution\cite{Neights2025}. Although these hypotheses present compelling theoretical possibilities, they face challenges in accounting for the full suite of observational features, particularly the luminous non-thermal X-ray emission detected approximately one day prior to the gamma-ray flares, significant spectral softening, and the off-center location within the host galaxy. Given these observational constraints, the jetted WD-IMBH TDE offers a more self-consistent and natural explanation for the observed properties of EP250702a.

The successful launch of a relativistic jet with an initial Lorentze factor of $\gtrsim 56$ (Supplementary material, constraints on the jet bulk Lorentz factor), implies a high spin of the black hole residing in \src{}\cite{Blandford1977, McKinney2012}. The dimensionless black hole spin parameter $a$ is estimated to be around 0.6 at a redshift of 1.036 (Fig.~\ref{fig:BH_spin} online; Supplementary material, constraints on the black hole spin). Either the IMBH was born with a high spin, or the IMBH was born as a stellar-mass black hole and grew via coherent accretion, while a growth history with chaotic accretion is disfavored\cite{Berti2008}. Based on the luminosity of EP250702a, we estimate the event rate density of similar events to be $\sim 10^{-13} $ $\rm Mpc^{-3}\,yr^{-1}$ (Supplementary material, event rate density), which is orders of magnitude lower than what is expected for normal IMBH-TDEs and SMBH jetted TDEs. This low rate finds a natural explanation in the jetted TDE scenario through relativistic beaming, which limits detection to only a small fraction of events directed toward us. Future observations of sources analogous to EP250702a can provide further insights into the formation and evolution of IMBHs, the launch of relativistic jets in extremely high accretion systems, as well as their surrounding stellar environments. Perhaps of greatest interest, WD-TDEs are prime sources of simultaneous electromagnetic and gravitational wave (GW) emission in the sensitivity regime of future GW interferometers, allowing for important cosmological applications\footnote{LISA mission webpage, \url{https://www.lisamission.org/}}. 

\clearpage


\captionsetup[table]{name={\bf Table}}
\captionsetup[figure]{name={\bf Fig.}}

\clearpage

\clearpage

\clearpage

\clearpage

\begin{addendum}

\item[Conflict Interests] The authors declare that they have no conflict of interest.

\item[Acknowledgments] This work is based on the data obtained with Einstein Probe, a space mission supported by the Strategic Priority Program on Space Science of Chinese Academy of Sciences, in collaboration with the European Space Agency, the Max-Planck-Institute for extraterrestrial Physics (Germany), and the Centre National d'Études Spatiales (France). We acknowledge the support from the National Natural Science Foundation of China (12333005, 12333004, 12573019), the National Key R\&D Program of China (2025YFF0511100), and the Strategic Priority Research Program of the Chinese Academy of Sciences (XDB0550200). J.Y. acknowledges the support from the National Natural Science Foundation of China (13001106). J.-H.C. and L.-X.D. acknowledge the support from the Natural Science Foundation of China and the Hong Kong Research Grants Council (N\_HKU782/23, HKU17314822, HKU17305523). C.J. acknowledges the support by the National Natural Science Foundation of China (12473016, L2324211). H.S. acknowledges the support from the Young Elite Scientists Sponsorship Program by China Association for Science and Technology (YESS20240218). J.-H.C. acknowledges the support from the National Natural Science Foundation of China (12503053). W.-H.L. acknowledges the support from the National Natural Science Foundation of China (12473012 and 12533005). Y.-H.Y. acknowledges the supported by the European Research Council through the Consolidator grant BHianca (grant agreement ID~101002761). YQX acknowledges the support from the National Natural Science Foundation of China (12025303). Y.-F.H. is supported by the National Natural Science Foundation of China (12233002), and by the National Key R\&D Program of China (2021YFA0718500). Y.-F.H. also acknowledges the support from the Xinjiang Tianchi Program. S.K.J. and R.L. acknowledge the support of the Anusandhan National Research Foundation through the grant CRG/2002/008253.  SKJ and RL thank Arvind Balasubramanian for useful discussions during the GMRT data analysis. We thank the staff of the GMRT for making these observations possible. This work is also based on observations made with the Gran Telescopio Canarias (GTC), installed at the Spanish Observatorio del Roque de los Muchachos of the Instituto de Astrofísica de Canarias, on the island of La Palma (under programme GTCMULTIPLE2G-25A; PI: F. Coti Zelati). F.C.Z. is supported by a Ramón y Cajal fellowship (grant agreement RYC2021-030888-I). F.C.Z., A.M., Y.-L.W. and N.R. acknowledge support from the Spanish grant ID2023-153099NA-I00 and from the programme Unidad de Excelencia María de Maeztu CEX2020-001058-M. GMRT is run by the National Centre for Radio Astrophysics of the Tata Institute of Fundamental Research. E.T. acknowledges support by the European Research Council through the Consolidator grant BHianca (grant agreement ID~101002761). S.K. acknowledges support from the President's International Fellowship Initiative of the Chinese Academy of Sciences. We acknowledge the National Astronomical Science Data Center of China (NADC) for their contributions to the development of the EP time domain astronomical information center platform, including the observation data management, AI based transient identification agent, which supported the data processing and analysis in this work.

\item[Author Contributions] Weimin Yuan has been leading the Einstein Probe project as Principal Investigator since the mission proposal stage. Chichuan Jin, Lixin Dai, Weimin Yuan, Bing Zhang, Dongyue Li, Wenda Zhang, and Huaqing Cheng initiated the study. Chichuan Jin, Lixin Dai, Bing Zhang, Dongyue Li, Wenda Zhang, and Weimin Yuan coordinated the scientific investigations of the event and led the subsequent discussions. Bing Zhang, Lixin Dai, Rong-Feng Shen, Xinwen Shu, Weimin Yuan, Bifang Liu, Hongyan Zhou, Dongyue Li, Wenda Zhang, and Jun Yang contributed to the theoretical investigation of the event. Dongyue Li and Huaqing Cheng processed and analyzed the EP-WXT data. Dongyue Li performed the EP-WXT data backward-stacking and trigger time identification. Dongyue Li processed and analyzed the EP-FXT data. Dongyue Li, Huaqing Cheng, Chichuan Jin, and Lixin Dai obtained the Chandra observation data. Dongyue Li reduced the Chandra observation data. Jun Yang processed and analyzed the Fermi/GBM data. Jun Yang performed Fermi/GBM data targeted backward search and positional verification. Huaqing Cheng, Wenjie Zhang, and Haiwu Pan performed the Fourier power spectral density and Lomb–Scargle periodogram analyses of the EP-FXT data. Jun Yang performed the background modelling and duration calculation of Fermi/GBM data. Yu-Han Yang and Jun Yang calculated the the minimum variability timescale of Fermi/GBM data. Jun Yang and Wenda Zhang constrained the black hole mass based on the minimum variability timescale. Dongyue Li performed the soft X-ray spectral analysis of the EP-WXT, EP-FXT and Chandra data. Dongyue Li contributed to the jet and disk spectral modelling. Jun Yang performed the joint soft X-ray and gamma-ray spectral analysis. Jun Yang contributed to constraining on the jet bulk Lorentz factor. Weihua Lei and Chang Zhou contributed to constraining on the black hole spin. Ning Jiang and Jiazheng Zhu contributed to the analysis on the host-galaxy properties. Fan Xu and Xinwen Shu performed the afterglow modelling. Jin-Hong Chen and Rong-Feng Shen performed the theoretical modelling of IMBH-WD TDE. Hui Sun contributed to the estimation of the event-rate density. Zhongnan Dong, Chun Chen, Xia Li, and Rong-Feng Shen contributed to the near-infrared data taking with SYSU-80cm. 
Francesco Coti Zelati, Luis Galbany, Maria Cristina Baglio, Alessio Marino, Yilong Wang, and Nanda Rea contributed to the data acquisition and analysis with GTC. Brenna Mockler, Christopher R. Burns, and Daniel Kelson contributed to the NIR data taking with Magellan. Stéphane Corbel, Alexis Coleiro, Noa Grollimund, Floriane Cangemi, Jérome Rodriguez, and Xinwen Shu contributed to the radio data taking with MeerKAT. Sitha K. Jagan, L. Resmi, and Sourya R. Das contributed to the radio data taking with GMRT. Huaqing Cheng, Yijia Zhang, Chang Zhou, Yehao Cheng, and Guoying Zhan are the transient advocates, and Zhixing Ling is the duty scientist on 3 July 2025 and contributed to the discovery and preliminary analysis of this event. Zhixing Ling, Chen Zhang, Xiaojin Sun, Shengli Sun, Yonghe Zhang, Zhiming Cai and Weimin Yuan contributed to the development of the WXT instrument. Chen Zhang, Zhixing Ling, Huaqing Cheng, and Yuan Liu contributed to the calibration of WXT data. Yuan Liu, Huaqing Cheng, Chichuan Jin, Wenda Zhang, Dongyue Li, Jingwei Hu, Heyang Liu, and Haiwu Pan contributed to the development of WXT data analysis software. Yong Chen and Shumei Jia contributed to the development of the FXT instrument and the development of FXT data analysis software. Dongyue Li, Wenda Zhang, Jun Yang, and Weimin Yuan drafted the manuscript with the help from all authors.

\end{addendum}

\clearpage

\include{sup}

\end{document}

%% file: sup.tex
\section*{\centering \large Supplementary material}

\subsection{Observations and data reduction.\\} \label{sec:obs}

Logs of the observations used in this work are presented in Tables~\ref{tab:xray_label}--\ref{tab:radio_obs}.

\noindent\textbf{EP-WXT.} The Einstein Probe (EP)\cite{Yuan2022}, launched on 9 January 2024, is a space mission led by the Chinese Academy of Sciences (CAS), in collaboration with the European Space Agency (ESA), the Max Planck Institute for Extraterrestrial Physics (MPE), Germany, and the France Space Agency (CNES). EP is an interdisciplinary X-ray observatory dedicated to time-domain astronomy and high-energy astrophysics. One of the two main payloads onboard EP is the Wide-field X-ray Telescope (WXT), which employs novel lobster-eye micro-pore optics (MPO) to achieve both a large instantaneous FoV of 3,600 square degrees and high sensitivity of $\sim(2\text{--}3)\times10^{-11}$\,$\rm erg\,cm^{-2}\,s^{-1}$ in the 0.5--4 keV band for an exposure time of 1\,ks. WXT data reduction was performed following the standard data reduction procedures implemented in the WXT Data Analysis Software (WXTDAS, Liu et al. in prep.), along with the calibration database (CALDB) maintained by the Einstein Probe Science Center (EPSC). The CALDB is constructed based on results from both on-ground calibration experiments\cite{2025ChengEPWXT} and in-orbit calibration campaign (Cheng et al. in prep.). Raw data are processed using the WXT data analysis chain tool \texttt{wxtpipeline}.

EP250702a was first detected by EP-WXT during an observation commencing at 2025-07-02T02:53:44 (UTC), about 10 hours before the Fermi/GBM trigger of GRB 250702D at 2025-07-02T13:09:02.03 (UTC)\cite{GCN40886}. The signal from the source was continuously detected in seven subsequent WXT observations. Among the total of eight WXT observations, the third, fourth, and fifth epochs temporally overlapped with GRB 250702D, B, and E, respectively (Fig.~\ref{fig:xray_evol} and Fig.~\ref{fig:xray_gamma}). The WXT localization of the source is R.A. = 284.700 deg and Dec. = -7.869 deg (J2000) with an uncertainty of 2.4 arcmin in radius at 90\% confidence level (including both statistical and systematic errors), spatially coincident with the gamma-ray sources\cite{GCN40906}. 

\noindent\textbf{EP-WXT data backward-stacking and trigger time identification.} The position of EP250702a was covered by EP-WXT more than 1,500 times before the first WXT detection. Stacking the pre-detection WXT data using WXTDAS task \texttt{wxtmerge} revealed that the source signal had already emerged one day earlier, on 1 July 2025. To determine the precise onset time of EP250702a's signal in the EP-WXT data, we progressively backward-stacked the earlier EP-WXT data and calculated the detection significance at the source position using the Li-Ma formula\cite{Li1983} every time a new observation is added into the stack. The trigger time was determined as the start time of the last data stack before the detection significance began to decrease. Through this method, we determined the trigger time of EP250702a in the EP-WXT data to be 2025-07-01T01:40:26 (UTC), more than one day before the Fermi/GBM trigger time of the repeating gamma-ray flares. Further stacking of earlier data yielded no significant detections, with the signal-to-noise ratio (SNR) calculated using the Li–Ma formula remaining below 4, resulting in a 0.3--10 keV flux upper limit at 90\% confidence level of $6.03 \times 10^{-12}~{\rm erg~cm^{-2}~s^{-1}}$ on 30 June 2025, assuming the same spectral shape as observed on 1 July 2025. This suggests that the X-ray brightness increased by a factor of more than 5 within one day.

\noindent\textbf{EP-FXT.} Follow-up observations of EP250702a in the soft X-ray band were carried out using the Follow-up X-ray Telescope (FXT)\cite{Chen2020SPIE} onboard EP. The FXT is another main payload onboard EP, operating in the 0.3--10 keV energy range. It comprises two co-aligned modules, FXT-A and FXT-B, each equipped with 54 nested Wolter-I paraboloid-hyperboloid mirror shells. The raw FXT data were processed using \texttt{fxtchain}, the standard data reduction pipeline included in the FXT Data Analysis Software (FXTDAS, version 1.20)\cite{2025ZhaoFXTDAS}. The procedure includes particle event identification, pulse invariant conversion, grade calculation and selection (grades 0--12), bad and hot pixel flagging, and the selection of good time intervals based on housekeeping files. The pipeline finally screens the raw data to produce cleaned event files and generates the exposure map and the response files. The light curve and energy spectrum of EP250702a were extracted from the cleaned event files using a circular region with a radius of 60$^{\prime\prime}$ centered on the source. For the background products, a circular region with a radius of 140$^{\prime\prime}$ covering a nearby source-free area was used.

\noindent\textbf{Chandra.} A {\it Chandra} Director's Discretionary Time observation of EP250702a (ID: 31003, PI: Dongyue Li) was conducted on 18 July 2025 using the Advanced CCD Imaging Spectrometer (ACIS-I), with an exposure time of 14.9 ks. The main objective of this observation was to obtain the most accurate X-ray localization of EP250702a and to constrain its soft X-ray spectrum and flux. Data reduction was performed following the standard procedures using the {\it Chandra} Interactive Analysis of Observations (CIAO, version 4.17.0) software\cite{Fruscione2006}. The data retrieved from the archive was reprocessed using the CIAO task \texttt{chandra\_repro} to apply the latest calibrations version 4.12.2.
Point-source detection was performed by utilizing the CIAO task \texttt{wavdetect} in 0.5--7.0\,keV (the default broad band for ACIS data) while using scales of 1, 2, 4, 8, and 16 pixels.
Only one X-ray source was detected within the EP-FXT error circle, thus providing the most accurate localization of EP250702a: R.A. = $18{\rm h}58{\rm m}45.56{\rm s}$, Dec. = $-7^{\circ}52^{\prime}26.1^{\prime\prime}$ (J2000), with a $1\sigma$ uncertainty of $0.79^{\prime\prime}$ that includes both statistical and systematic errors. To enhance the positional accuracy, we attempted absolute astrometric corrections. However, since most X-ray sources in the field lack clear optical counterparts, it was not feasible to apply positional refinements. Therefore the positions provided in this work are directly taken from \texttt{wavdetect}, with the associated uncertainties incorporating the systematic errors. The ACIS-I spectrum and response files were generated using the CIAO task \texttt{specextract}. The source photons were extracted from a circular region with a radius of 5$^{\prime\prime}$ centered on the source, while the background photons were extracted from an annulus region centered on the same position with inner and outer radii of 10$^{\prime\prime}$ and 20$^{\prime\prime}$, respectively.

\noindent\textbf{Fermi/GBM.} Fermi/GBM detected and localized three gamma-ray flares on 2 July 2025, designated GRB 250702D, B, and E (hereafter GRB 250702B), with trigger times at 2025-07-02T13:09:02 ($T_{\rm 0, D}$), 2025-07-02T13:56:05 ($T_{\rm 0, B}$), and 2025-07-02T16:21:33 ($T_{\rm 0, E}$), respectively\cite{GCN40883, GCN40886, GCN40890}. These three bursts likely originated from the same astrophysical source and exhibited both spatial and temporal coincidence with the soft X-ray transient EP250702a detected by EP\cite{GCN40891, GCN40906, GCN40931}. To analyze the gamma-ray data of GRB 250702B, we first retrieved the daily time-tagged event dataset covering the relevant time intervals from the Fermi/GBM public data archive. The analysis procedure follows the methodology described in Refs.\cite{Yang2020ApJ, Yang2022Natur}. For all three bursts, detectors nb and b1 represent the sodium iodide (NaI) and bismuth germanium oxide (BGO) detectors with the smallest viewing angles in respect to the localization of EP250702a, respectively. Therefore, we selected these two detectors for the subsequent temporal and spectral analyses.

\noindent\textbf{Fermi/GBM data targeted backward search and positional verification.} Motivated by the detection of the EP250702a signal on 1 July 2025 through backward stacking of EP-WXT data, we conducted a targeted backward search for EP250702a in the Fermi/GBM dataset over a two-day period preceding Flare D. With the aid of visual inspection, this search revealed a candidate weak flare (hereafter Flare A) at 2025-07-01T11:55:18 ($T_{\rm 0, A}$), which may be associated with EP250702a. We applied the Bayesian block algorithm\cite{Scargle2013ApJ} to Flare A and calculated the signal-to-noise ratio for each block based on the method described in Ref.\cite{Vianello2018ApJS}. The block with the highest signal-to-noise ratio, lasting 6 seconds, reaches a significance level of 8$\sigma$. To further evaluate the positional association between Flare A and EP250702a, we performed localization analysis for Flares A, D, B, and E using the BALROG method\cite{Burgess2018MNRAS, Berlato2019ApJ}, which simultaneously fits the spectral and spatial properties of the event data of GBM detectors. The resulting localizations were then compared with the soft X-ray position of EP250702a. Following the guidelines in Ref.\cite{Berlato2019ApJ}, we selected a 10-second time interval centered on a single peak in the light curve and used detectors from n6 to nb together with b1 (the b1 side of the spacecraft) for each flare in this analysis. As shown in Fig.~\ref{fig:gbm_balrog} and Table~\ref{tab:gbm_balrog}, the localization results for Flares A, D, B, and E are generally consistent with each other and broadly consistent with the soft X-ray position of EP250702a. Specifically, the soft X-ray position of EP250702a falls within the 2$\sigma$ statistical localization contours of Flares A and D, and within the 1$\sigma$ contours of Flares B and E, providing statistical support for the interpretation that Flare A likely originated from EP250702a as the earlier gamma-ray flare. However, given that the source lies close to the Galactic center, we cannot rule out the possibility that this candidate flare is unrelated.

\noindent\textbf{SYSU-80cm.} We observed the field of EP250702a using the Sun Yat-sen University 80cm infrared telescope with $58\times20$\,s exposures in the $J$ band (1.17--1.33\,${\rm\mu m}$)\cite{GCN40986}. The observations began at 2025-07-03T19:36:00 (UTC), 1.27\,days after the Fermi/GBM trigger of GRB 250702D. We did not detect any near-infrared counterpart at the position of the afterglow, down to a 5$\sigma$ depth of 17.5\,mag (Vega system) in $J$ band.

\noindent\textbf{GTC.} We observed EP250702a on 3 July 2025 using the Optical System for Imaging and low-Intermediate-Resolution Integrated Spectroscopy (OSIRIS$+$\cite{Cepa2003}) mounted on the 10.4-m Gran Telescopio Canarias (GTC). The observations were carried out under clear skies and good seeing conditions, with an average seeing of $\simeq0.7^{\prime\prime}$. A sequence of four 60-s images was acquired in the Sloan $i$ filter, using a detector binning of 2$\times$2. To enhance sensitivity, the images were co-added using standard routines in IRAF\footnote{\url{https://iraf-community.github.io/}} software\cite{iraf1986, iraf1993}. An astrometric solution was then derived by matching field stars to the 2MASS\footnote{\url{https://irsa.ipac.caltech.edu/Missions/2mass.html}} reference catalog\cite{2mass} using the Starlink\footnote{\url{http://starlink.eao.hawaii.edu/starlink}} GAIA tool\cite{starlink}. Aperture photometry was performed using \texttt{daophot}\cite{Stetson1987} on the image after World Coordinate System calibration, adopting an aperture radius of $\simeq$ 1.5 times the full width at half maximum of the stellar point spread function. Photometric calibration was obtained using eight non-saturated reference stars in the field with Sloan $i$ magnitudes tabulated in the Pan-STARRS catalog\cite{Chambers2016}. No source was detected at the position of the infrared counterpart reported by Ref.\cite{Levan2025}. A 3$\sigma$ upper limit on the magnitude was thus estimated by computing the root-mean-square (RMS) of background-subtracted fluxes in blank apertures with the same geometry, after masking detected sources. This yields $i>22$ (AB system) for the optical counterpart of EP250702a.

\noindent\textbf{MeerKAT.} We performed multi-epoch radio observations of EP250702a with MeerKAT using the L-band (856--1,712 MHz) and S4-band (2,625--3,500 MHz) receivers between 7 July 2025 and 21 July 2025 (ID: MKT-24172, PI: Corbel). We used PKS J1939-6342 as a primary calibrator to set the absolute flux and bandpass, and J1822-0938 (resp. J1833-2103) at S band (resp. L-band) for complex gain calibration. Each run consisted of a single 11-minute scan of EP250702a, interleaved by 1.75-minute scans of the phase calibrator, as well as a single 5-minute scan of the primary calibrator. The correlator was configured to deliver 4,096 channels across the total bandwidth, which were binned down to 1,024 channels, with an 8-second integration time per visibility point. We reduced and imaged the data using the oxkat data reduction scripts\cite{oxkat}. The visibilities were initially flagged and calibrated using the Common Astronomy Software Application (CASA) package\cite{casa}. Additional flagging on the target data was conducted using the Tricolour package\cite{tricolour}. For imaging the field of EP250702a, we employed WSCLEAN\cite{wsclean} with a Briggs weighting scheme (robust parameter of $-0.3$). We generated full-band, multi-frequency synthesis (MFS) images by deconvolving in eight sub-bands. After a step of direction-independent self-calibration with the CubiCal package\cite{cubical}, the target was imaged a second time using masked deconvolution. This reduction process yielded radio maps with an angular resolution of $\sim 3^{\prime\prime}$ (resp. $\sim 6^{\prime\prime}$) and an average RMS noise level of $14~\mu\text{Jy}$ (resp. $22~\mu\text{Jy}$) at S-band (resp. L-band). To extract the positions and flux densities, we fitted the source in the image plane with an elliptical Gaussian model. The best radio position was obtained by averaging the fitted source positions over all S-band observations, giving R.A. = $18{\rm h}58{\rm m}45.555{\rm s}$ and Dec. = $-7^{\circ}52^{\prime}26.49^{\prime\prime}$ (J2000), with a 1$\sigma$ error of $0.3^{\prime\prime}$.

\noindent\textbf{GMRT.} We carried out observations of the radio counterpart of EP250702a with the upgraded Giant Metrewave Radio Telescope (uGMRT), as part of the Target of Opportunity program 48\_167 (PI: Resmi L). We conducted Band-5 (1--1.46\,GHz) observations on 8 July 2025, 7.7\,days after the EP-WXT trigger time, obtaining a 3$\sigma$ upper limit of 263\,$\mu$Jy at the source position from a 1-hour observation. A 4-hour follow-up observation in the same band on 12 July 2025 (11.8\,days post-trigger) led to the detection of the source with a flux density of 82.3\,$\mu$Jy at 1.26\,GHz. We also conducted additional observations in Band-4 (0.55--0.85\,GHz) on 30 July 2025 (29.7\,days after the trigger) and measured a flux density of 111.8\,$\mu$Jy at 0.65\,GHz. We continued monitoring the source throughout uGMRT Cycle 48, and the results are presented in Table~\ref{tab:radio_obs}. The data were processed using the CAPTURE-CASA6 pipeline\cite{Kale2021ExA}, applying standard flagging and calibration procedures, with 3C 48 used as the primary flux calibrator and 1911-201 as the phase calibrator. Flux densities were derived using a two-dimensional Gaussian fit at the source position on the final images by employing the CASA task \texttt{imfit}\cite{casa}.

\subsection{Timing analysis.\\}

\noindent\textbf{Soft X-rays.} The EP-WXT and early-time EP-FXT light curves exhibit pronounced short-timescale variability. To investigate the late-time timing properties of EP250702a in the X-ray band, we performed Fourier power spectral density (PSD) analysis for each EP-FXT observation using the stingray package\cite{Huppenkothen2019}, adopting a timing resolution of 0.1\,s and a segment length of 512\,s. The PSDs were normalized according to the method of Ref.\cite{Leahy1983}, such that the expected power from pure Poisson noise is equal to 2. For the majority of the observations, the PSDs are dominated by Poisson noise across nearly all measured frequencies, due primarily to the limited source count rate. In a few observations with relatively high count rates, we detect excess red noise at low frequencies; however, no significant periodic or quasi-periodic signals are observed. As an illustrative example, the PSD of the first EP-FXT observation, which has the highest source count rate among all EP-FXT observations, is shown in Fig.~\ref{fig:fxt_obs1_psd}a. To improve the signal-to-noise ratio, data from the FXT-A and FXT-B modules were combined. The best-fit power-law model is also plotted, from which we simulated 100,000 light curves based on the method of Ref.\cite{TimmerKonig1995}. The PSDs of these light curves were calculated, leading to the determination of the distribution of the variability power at each Fourier frequency. The $3\sigma$ confidence level of the power is plotted for illustration. We thus conclude that the significance of any potential QPO signal, if present, is below $3\sigma$. The stacked PSD from all observations operated in imaging mode (FF+PW) is calculated and presented in Fig.~\ref{fig:fxt_obs1_psd}b, which is found to be dominated by white noise at all frequencies. To further search for potential periodic signals, we calculated the Lomb–Scargle periodogram using data from the first EP-FXT observation. However, no significant periodicity was identified within the frequency range of $\sim 2 \times 10^{-3}$--$10$ Hz. This result is consistent with the findings from a NuSTAR observation conducted approximately two days after the EP trigger time\cite{GCN41014}.

\noindent\textbf{Gamma-rays.} Accurate background modelling is essential for the timing analysis of the Fermi/GBM data for GRB 250702B. The background modelling procedure began with the application of the baseline correction method\cite{pybaselines} and the Bayesian block algorithm\cite{Scargle2013ApJ} to the observed light curves, which provided initial background estimates and identified pulse structures, respectively. We then calculated the SNR for each Bayesian block, assuming Poisson-distributed data with a Gaussian background\cite{Vianello2018ApJS}. A polynomial model was subsequently fitted to the blocks with SNR greater than 3 to construct the final background model. 
The background-subtracted light curves of GRB 250702B, derived from the data of nb and b1 detectors and binned at 10-second bin size, are shown in Fig.~\ref{fig:xray_gamma}a. 

As illustrated in Fig.~\ref{fig:xray_gamma}a, both Flares A and D are characterized by a single dominant pulse overlaid with several sharp spikes, with durations of $T_{\rm 90, A} = 20_{-1}^{+2}~{\rm s}$ and $T_{\rm 90, D} = 82_{-12}^{+4}~{\rm s}$, respectively. Flare B similarly features a main pulse but is accompanied by several weaker pulses occurring both before and after the primary emission, resulting in a total duration of $T_{\rm 90, B} = 426_{-10}^{+11}~{\rm s}$. Flare E exhibits two prominent pulses, with durations of $T_{\rm 90, E1} = 97_{-11}^{+25}~{\rm s}$ and $T_{\rm 90, E2} = 160_{-16}^{+9}~{\rm s}$, separated by a quiescent interval of approximately 200 seconds dominated by background. It is worth noting that the b1 light curves of GRB 250702B show pronounced pulse structures, suggesting substantial high-energy photon emission. This high-energy emission is further supported by the spectral analysis results presented below. A comparison with the simultaneous soft X-ray light curves detected by EP, also shown in Fig.~\ref{fig:xray_gamma}a, reveals that both gamma-ray and soft X-ray light curves share similar temporal profiles.

The minimum variability timescale (MVT, denoted as $\delta t_{\rm min}$) refers to the shortest timescale over which a statistically significant change in count rate can be observed in the light curve. This temporal feature is closely associated with the geometric scale of the central engine and the emission region\cite{Bhat1992Natur, Kobayashi1997ApJ, Golkhou2014ApJ, Camisasca2023AA}. To estimate the MVTs of GRB 250702B, we applied the structure function method based on the Haar wavelet transform, as introduced in Refs.\cite{Golkhou2014ApJ, Golkhou2015ApJ}, to the 10-ms resolution background-subtracted light curves. The results are summarized in Table~\ref{tab:x_gamma_prop}. We note that the derived MVTs are generally on the order of a few seconds. As illustrative examples, the two shortest MVT results are presented in Fig.~\ref{fig:gbm_mvt}. Taking the minimum value $\delta t_{\rm min} \approx 1.5~{\rm s}$, corresponding to $\delta t_{\rm min}/(1+z) \approx 0.74~{\rm s}$ in the source rest frame, the most stringent upper limit on the characteristic size of the source emitting region can be constrained as $R_{\rm E} < c\delta t_{\rm min} / (1 + z) \approx 2.2\times10^{10}~{\rm cm}$. Given that the size cannot be smaller than the black hole event horizon, we can put an upper limit on the black hole mass of $M_{\rm BH} \lesssim R_{\rm E} c^2/2{\rm G} < 7.5\times10^4~{M_\odot}$ by assuming a Schwarzschild black hole, where $\rm G$ is the gravitational constant. This constraint suggests the presence of an intermediate-mass black hole in EP250702a.

Such sub-second variability can arise from mini-jets within a Poynting-flux-dominated outflow. This mechanism has been proposed in the Internal-Collision-induced Magnetic Reconnection and Turbulence (ICMART) model for GRBs\cite{Zhang2011ApJ} and in the jets-in-a-jet scenario developed for blazars\cite{Giannios2009MNRAS}. Since mini-jets can produce variability timescales shorter than those associated with the central engine, the constraint on the black hole mass described above becomes less stringent, allowing for a larger inferred black hole mass.

\subsection{Spectral analysis.\\}

\noindent\textbf{Soft X-rays.} We performed soft X-ray spectral analysis for the EP-WXT, EP-FXT, and {\it Chandra} observations using Xspec (version 12.13.0c)\cite{Arnaud1996ASPC}. Throughout the spectral fitting process, the elemental abundances were set to \textit{WILM}\cite{Wilms2000}, and the photoelectric cross-sections were set to \textit{VERN}\cite{Verner1996}. 

Due to limited photon statistics, the EP-WXT spectra were grouped with a minimum of two counts per bin in 0.5--4\,keV energy band, and the best-fit parameters were obtained by minimizing the Cash statistic\cite{Cash1979ApJ}. Spectral modelling was carried out using a simple absorbed power-law model \textit{tbabs*ztbabs*powerlaw} in Xspec, where \textit{tbabs} contributes to the Galactic absorption with a hydrogen column density $N_{\rm H}=3.3 \times 10^{21}~{\rm cm^{-2}}$, and \textit{ztbabs} contributes to the intrinsic abosorption, with the redshift fixed at 1.036. It is found that the absorbed power-law model provides statistically acceptable fits for all WXT observations, with the fitting results summarized in Table~\ref{tab:xray_label}. All uncertainties are reported at the $1\sigma$ confidence level for a single parameter of interest unless otherwise stated. 

For the EP-FXT observations, spectra were grouped with a minimum of 25 counts per bin when the total counts exceeded 625 in the 0.5–10\,keV energy band, and the $\chi^2$ statistic was used for spectral fitting. Otherwise, for spectra with total counts between 100 and 500, minimum grouping value of 10 was applied, while spectra with fewer than 100 total counts were grouped with a minimum of 1 count per bin. In both latter cases, the Cash statistic\cite{Cash1979ApJ} was adopted for spectral fitting. The FXT spectra were initially fitted with a simple absorbed power-law model, \textit{const*tbabs*ztbabs*powerlaw}, where the \textit{const} component was applied to correct for the calibration differences between the FXT-A and FXT-B units, with that for FXT-B fixed at unity as a reference. This model provides statistically acceptable fits for almost all FXT datasets taken within about 2 weeks after the trigger. The spectral fitting indicated a change in the absorption properties. An intrinsic hydrogen column density, beyond the Galactic component, was required for the first nine FXT observations (until 10 July, Obs ID: 06800000743). Subsequently, beginning with the eleventh observation on the late evening of 10 July (eight days after the trigger), no additional absorption was necessary. The best fitting parameters and the corresponding 1$\sigma$ uncertainties for the FXT observations are listed in Table~\ref{tab:xray_label}. The spectra exhibited softening over the monitoring period, with the photon index increasing from approximately 1 to greater than 2, as illustrated in Fig.~\ref{fig:xray_evol}b.

For the {\it Chandra} observation taken on 18 July 2025, approximately 40 net counts were collected during the 14.9 ks exposure in the 0.5--7 keV energy band. The spectra was grouped with a minimum of three counts per bin. A statistically acceptable fit was obtained using a simple absorbed power-law model, with the hydrogen column density fixed at the Galactic value and a photon index of $2.6 \pm 0.6$. However, significant structures are evident in the residuals, suggesting that the simple absorbed power-law model does not yield an ideal fit. Due to the limited photon statistics, spectral fitting with more complex models was not feasible using the {\it Chandra} data alone. To improve the spectral constraints, we therefore performed a joint fit combining the simultaneous {\it Chandra} and EP-FXT data.

\noindent\textbf{Joint spectral fitting reveals two distinct components.} We conducted joint spectral fitting of the simultaneous EP-FXT and Chandra spectra. The initial model \textit{const*tbabs*ztbabs*powerlaw} indicated negligible intrinsic absorption beyond the Galactic component. However, the fit was unsatisfactory, as the residuals revealed significant structural features around 2\,keV, as shown in Fig.~\ref{fig:joint_spec}(a). To improve the model, we added a second power-law component, using \textit{const*tbabs*} \textit{(powerlaw+powerlaw)}. This two-component model resulted in a significantly better fit. An F-test confirmed the statistical necessity of the additional component, yielding a p-value of 0.02. The photon indices of the two power-law components were found to be $\Gamma_{1}=0.7^{+1.2}_{-2.3}$ and  $\Gamma_{2}=4.75^{+1.5}_{-1.0}$. 
By using Bayesian inference, we also obtained consistent constraints on the photon indices of the two power-law components. The posterior probability distributions, shown in Fig.~\ref{fig:joint_spec}b, clearly reveal two distinct photon indices of $\Gamma_{1}=0.7_{-1.2}^{+2.0}$ and $\Gamma_{2}=4.5_{-0.9}^{+2.3}$, which do not overlap at the 1$\sigma$ confidence intervals. Furthermore, a comparison of the Bayesian evidence for the double power-law model versus the single power-law model yields a Bayes factor of log$_{10}(\mathcal{Z}_{\rm pl+pl}/\mathcal{Z}_{\rm pl})=0.74$, providing substantial support for the double power-law model\cite{kass1995}.
This provides clear and direct evidence for the presence of two distinct spectral components. The inferred photon index of the second power-law component is exceptionally soft (with a photon index of $\sim5$), which is naturally explained by thermal emission from a warm photosphere or an accretion disk. Therefore, motivated by this theoretical consideration, we replaced the second power-law with a physical blackbody component, yielding the final model \textit{const*tbabs*(bbody+powerlaw)}. We note that since the \textit{const*tbabs*(powerlaw+powerlaw)} and \textit{const*tbabs*(bbody+powerlaw)} models provide statistically equivalent fits, neither can be ruled out solely based on fit quality. However, we favor the latter model, as it aligns with the physical expectation of thermal emission from a newly-formed accretion disk in a TDE. The best-fit parameters and statistical results for all three models are summarized in Table~\ref{tab:joint_fit_xray}.

\noindent\textbf{Simultaneous EP-WXT+Fermi/GBM spectral fits.} Thanks to the early, long-term wide-field coverage of EP to EP250702a, the broad-band data collected from EP-WXT and Fermi/GBM enable us to examine the spectral energy distribution (SED) from soft X-rays to gamma-rays. We identified four flaring intervals that were simultaneously detected by both missions, as well as two intervals observed exclusively by Fermi/GBM. The corresponding time ranges are listed in Table~\ref{tab:x_gamma_prop} and marked as SED1–6 in Fig.~\ref{fig:xray_gamma}. For the EP-WXT data, we extracted the source spectrum, background spectrum, and generated the corresponding detector redistribution matrix and ancillary response for each of the four time intervals. For the Fermi/GBM data, we followed the methodology described in Refs.\cite{Yang2022Natur, Yang2023ApJ} to construct the source and background spectra for each time interval and for each detector (nb and b1). The detector response matrix was produced using the gbm\_drm\_gen package\cite{Burgess2018MNRAS, Berlato2019ApJ}. 

We then perform spectral fitting using the Bayesian inference approach described in Refs.\cite{Yang2022Natur, Yang2023ApJ, Yin2025arXiv}. For the EP-WXT data, we employ the CSTAT\cite{Cash1979ApJ} statistic, which is appropriate for Poisson-distributed data with a Poisson background, whereas the PGSTAT\cite{Arnaud1996ASPC} statistic is used for the Fermi/GBM data, as it is suitable for Poisson-distributed data with a Gaussian background. To fit the joint spectra from soft X-rays to gamma-rays, the spectral model includes the following two components:
\begin{itemize}
    \item Absorption component: We adopt the Tuebingen-Boulder absorption models\cite{Wilms2000ApJ}, \textit{tbabs} and \textit{ztbabs}, to account for the Galactic and intrinsic absorption components, respectively. In our spectral fits, the $N_{\rm H}$ parameter of the \textit{tbabs} model is fixed at the Galactic value of $3.3 \times 10^{21}~{\rm cm^{-2}}$ (Ref.\cite{Willingale2013}). For the \textit{ztbabs} model, the redshift is set to $z = 1.036$\cite{gompertz2025jwst}, and the $N_{\rm H}$ parameter is fixed at $2.59 \times 10^{22}~{\rm cm^{-2}}$, as derived from the first FXT observation.
    \item Intrinsic spectral component: To describe the intrinsic emission from the source, we consider four empirical models: a power-law model (\textit{pl}), a cutoff power-law model (\textit{cpl}), the Band function\cite{Band1993ApJ} (\textit{band}), and a smoothly broken power-law model (\textit{sbpl}).
\end{itemize}

The spectral fitting results and corresponding fit statistics for the models described above are summarized in Table~\ref{tab:x_gamma_prop}. The model-predicted SEDs derived from the spectral fittings at different time intervals are displayed in Fig.~\ref{fig:xray_gamma}b. We note that our results are generally consistent with those reported in Refs.\cite{GCN40931, Oganesyan2025arXiv}. Furthermore, based on the Bayesian information criterion (BIC)\cite{Schwarz1978AnSta}, our joint spectral fits do not favor the simple power-law model. The best-fitting intrinsic model is \textit{cpl} for SED1, 2, 4, and 5, and \textit{sbpl} for SED3 and 6, indicating the presence of a spectral peak in the $\nu F_{\nu}$ spectrum. For these best-fit models, the low-energy spectral index $\alpha$ is around $-1$, and the peak energy $E_{\rm p}$ lies in the range of approximately 600--2000 keV, except for the earlier SED1, which has a lower peak energy of about 100 keV. Given that EP-WXT and Fermi/GBM provided only partial coverage of EP250702a, the total fluence during the flaring phase, inferred from spectral fitting, is $\gtrsim (1.7\pm0.1)\times10^{-4}~{\rm erg~cm^{-2}}$, corresponding to an isotropic-equivalent energy of $\gtrsim (5.0\pm0.3)\times10^{53}~{\rm erg}$ at a redshift of $z=1.036$.

\subsection{Constraints on the jet bulk Lorentz factor.\\}

Inside the jet, high-energy photons can be absorbed by low-energy photons through electron–positron pair production ($\gamma\gamma \rightarrow e^+e^-$) if the jet Lorentz factor $\Gamma_{\rm j}$ is not sufficiently large. When photon attenuation due to pair production becomes significant, the gamma-ray spectrum from the jet is expected to exhibit an exponential cutoff above a characteristic photon energy, $E_{\rm cut}$, at which the optical depth for pair production reaches unity, i.e., $\tau_{\gamma\gamma} (E_{\rm cut}) \sim 1$. This condition constrains a specific combination of the jet Lorentz factor $\Gamma_{\rm j}$ and the generally unknown emission radius $R$\cite{Gupta2008MNRAS, Zhang2009ApJ}. Within the internal shock model, however, $R$ can be related to the observed variability timescale $\delta t$ and $\Gamma_{\rm j}$ through $R \simeq 2\Gamma_{\rm j}^2 c \delta t$, which in turn enables us to place a constraint on $\Gamma_{\rm j}$\cite{Baring1997ApJ, Lithwick2001ApJ, Ravasio2024AA}. Following Ref.\cite{Ravasio2024AA}, their Eq.~(4) therein can be rewritten in the form of
\begin{equation}
    \tau_{\gamma\gamma}(E_{\rm cut}) = \frac{\eta(\beta) \sigma_{\rm T} d_{\rm L}^2 F_{\rm\nu}(E_{\rm th})}{2 c^2 \delta t \Gamma_{\rm j}^4},
\end{equation}
where $\eta(\beta)$ is a function of the photon spectral index $\beta$ that is given by $\eta(\beta) = (7/6)(1-\beta)^{-1}(-\beta)^{-5/3}$ (Ref.\cite{Svensson1987MNRAS}), $E_{\rm th}$ denotes the threshold energy above which target photons can interact with photons at $E_{\rm cut}$ to produce pairs, satisfying the relation of $E_{\rm th} E_{\rm cut} \sim 4 [\Gamma_{\rm j} / (1 + z)]^2 (m_{\rm e} c^2)^2$, and $F_{\rm\nu}(E_{\rm th})$ is the specific flux at $E_{\rm th}$. In general, both $E_{\rm th}$ and $E_{\rm cut}$ lie within the high-energy power-law regime characterized by the photon spectral index $\beta$. Therefore, $F_{\rm\nu}(E_{\rm th})$ can be related to $F_{\rm\nu}(E_{\rm p})$ through the scaling relation of $F_{\rm\nu}(E_{\rm th}) = F_{\rm\nu}(E_{\rm p}) (E_{\rm th} / E_{\rm p})^{\beta + 1}$. By solving $\tau_{\gamma\gamma} (E_{\rm cut}) \sim 1$, one can obtain an estimate of $\Gamma_{\rm j}$: 
\begin{equation}
    \Gamma_{\rm j} = \Bigg\{\frac{\eta(\beta) \sigma_{\rm T} d_{\rm L}^2 F_{\rm\nu}(E_{\rm p})}{2 c^2 \delta t} \bigg[\frac{(2 m_{\rm e} c^2)^2}{E_{\rm cut} E_{\rm p} (1 + z)^2}\bigg]^{\beta + 1}\Bigg\}^{\frac{1}{2 - 2\beta}}. \label{eq:gamma_estimate}
\end{equation}

To estimate the bulk Lorentz factor $\Gamma_{\rm j}$ of the jet based on Eq.~(\ref{eq:gamma_estimate}), one needs to input parameters obtained from spectral fitting, along with the minimum variability timescale and redshift, i.e., $\mathcal{P} = \{E_{\rm cut}, E_{\rm p}, \beta, F_{\rm\nu}(E_{\rm p}), \delta t, z\}$. During the flaring episodes of EP250702a, our joint WXT+GBM spectral fits do not reveal the presence of an exponential cutoff at high energies beyond the $\beta$ regime. It is worth noting that the exponential cutoff of \textit{cpl} model observed in SED1, 2, 4, and 5 is likely not physical, but rather a consequence of the detector's limited energy coverage or insufficient high-energy photon counts. In such cases, the high-energy photon spectral index in the \textit{band} or \textit{sbpl} model is typically difficult to constrain accurately, as reflected by the large uncertainties of $\beta$ in the Table~\ref{tab:x_gamma_prop}. Similar issues have been widely encountered during spectral fitting with instruments such as Swift/BAT and Fermi/GBM\cite{Lien2016ApJ, Gruber2014ApJS}. When the number of high-energy photons is sufficient, as in the case of SED3 and 6, the spectral data tend to favor a high-energy power-law component rather than an exponential cutoff. To set a lower limit on $E_{\rm cut}$, we introduce a cutoff feature at high energies into the best-fit \textit{sbpl} model of SED3 and 6. We then gradually shift $E_{\rm cut}$ toward lower energies until the BIC increases by more than 6 (i.e., $\Delta{\rm BIC} > 6$), which provides strong evidence against the model with cutoff\cite{kass1995}. We find that $\Delta{\rm BIC} > 6$ holds for any value of $E_{\rm cut}$, implying that $E_{\rm cut}$ should be larger than the highest energy covered by Fermi/GBM, i.e., $E_{\rm cut} \gtrsim 38~{\rm MeV}$. Together with $E_{\rm p}$, $\beta$, and $F_{\rm\nu}(E_{\rm p})$ obtained from the best-fit \textit{sbpl} model for SED3 and 6, as well as the minimum variability timescale $\delta t_{\rm min}$ for these intervals and the redshift of 1.036\cite{gompertz2025jwst}, the lower limit of $\Gamma_{\rm j}$ can be estimated as $\Gamma_{\rm j} \gtrsim 56$ for SED3 and $\Gamma_{\rm j} \gtrsim 60$ for SED6, consistent with values reported for previous jetted TDEs\cite{Burrows2011, Bloom2011, Andreoni2022}.

\subsection{Constraints on the black hole spin.\\}

We can constrain the IMBH spin by using the peak isotropic X-ray luminosity $L_{\rm X, iso}^{\rm peak} \sim 3\times 10^{49}~\rm erg~s^{-1}$ (taking $z=1.036$) if identifying the Blandford-Znajek (BZ) mechanism\cite{Blandford1977} as the jet launching mechanism, i.e., $\eta_{\rm BZ} L_{\rm BZ}=f_{\rm b} L_{\rm X, iso}$, where $\eta_{\rm BZ}$ is the efficiency of converting BZ power to X-ray radiation\cite{LeiZhang2011, Andreoni2022}. The BZ jet power $L_{\rm BZ}$ from a BH with mass $M_{\bullet}$ and 
angular momentum $J_{\bullet}$ is\cite{Lee2000, LiLX2000,WangDX2002,McKinney2005}
\begin{equation}
L_{\rm BZ}=1.7 \times 10^{42} a_{\bullet}^2 M_{\bullet,5}^2 
B_{\bullet,6}^2 F(a_{\bullet}) \ {\rm erg \ s^{-1}},
\label{eq_Lmag}
\end{equation}
where $a_{\bullet}=J_\bullet c/(G M_{\bullet}^2)$ is the BH spin parameter, 
$M_{\bullet,5}=M_{\bullet}/10^5~{M_{\odot}}$, 
$B_{\bullet,6}=B_{\bullet}/10^6~{\rm G}$ and $F(a_{\bullet})=[(1+q^2)/q^2][(q+1/q) \arctan q-1]$, here $q= a_{\bullet} /(1+\sqrt{1-a^2_{\bullet}})$. An isolated BH does not carry a magnetic field, $B_{\bullet}$ is closely related to the accretion rate $\dot M$ and the radius of the BH, which depends on $M_{\bullet}$. Combining these dependence, one finds that $L_{\rm BZ}$ is essentially a function of $\dot M$ and $a_{\bullet}$. {Since the BZ power is proportional to the accretion rate (i.e., $L_{\rm BZ} \propto \dot{M}$) and $L_{\rm X} \propto L_{\rm BZ}$ is assumed, one can normalize the accretion rate using the total accreted mass based on the observed flux and fluence of the event. The peak accretion rate can thus be estimated as $\dot M_{\rm peak} =M_* F_{\rm X}^{\rm peak}(1+z)/S_{\rm X} $, where $M_*$ is the WD mass}, $F_{\rm X}^{\rm peak}$ is the peak X-ray flux, and $S_{\rm X}$ is the total X-ray fluence (can be taking as a good proxy of the total mass of the tidally disrupted star).

The beaming factor $f_{\rm b}$ of the jet can be estimated with  $f_{\rm b}\sim \max(1/(2\Gamma_{\rm j}^2), \theta_{\rm j}^2/2)$, or with the event rates $f_{\rm b}\sim R_{\rm obs}/(10\% R_{\rm IMBH-TDE})$, where $\Gamma_{\rm j}$ and $\theta_{\rm j}$ are the Lorentz factor and the opening angle of jet. Using the Lorentz factor $\Gamma_{\rm j} \sim 56$ (Supplementary material, constraints on the jet bulk Lorentz factor, and taking $z=1.036$), we have $f_{\rm b} \sim 1.6 \times 10^{-4}$. It is hard to constrain the opening angle with the current data. We, therefore, give an estimation of $f_{\rm b} \sim 5\times 10^{-3}$ by adopting a typical opening angle $\theta_{\rm j} \sim 0.1$ rad (Supplementary material, afterglow modelling). The event rate density for EP250702a-like events is $R_{\rm obs} \sim 2.6^{+5.9}_{-2.1} \times 10^{-13} \rm Mpc^{-3}\,yr^{-1} $ (Supplementary material, event rate density). The event rate density for non-jetted IMBH-TDEs is $R_{\rm IMBH-TDE} \sim 4.4 \times 10^{-9} $ $\rm Mpc^{-3}\,yr^{-1}$ \cite{Jin2025}, one can find $f_{\rm b}\sim 5.9\times 10^{-4
}$. 

Now we can infer the BH spin $a_{\bullet}$ from EP X-ray observations. Considering the potentially high radiation efficiency of a magnetically powered jet, we adopt $\eta_{\rm BZ} \sim 0.5$ in the calculation. The constraint on $a_{\bullet}$ is presented in Fig.~\ref{fig:BH_spin}. For $M_* = 0.1~{M_\odot}$, the BH spin $a_{\bullet}$ is 0.82 (0.99, 0.93) for $f_{\rm b} \sim 1.6 \times 10^{-4}$ ($5\times 10^{-3}$, $5.9\times 10^{-4}$).

\subsection{Host-galaxy properties.\\} 

We obtained $H$- and $K_s$-band images of EP250702a using the FourStar near-infrared camera\cite{Persson2013FourStar} on the 6.5 m Magellan Baade telescope at the Las Campanas Observatory on 19 July 2025. Observations employed a standard dither pattern with 22 pointings per band, using short-exposure sequences of 8.833 s (13 loops) and 5.822 s (20 loops) for the $H$ and $K_s$ bands, respectively. All exposures were subsequently stacked during data reduction, resulting in total exposure times of 2,497.638 s and 2,561.68 s for $H$ and $K_s$ bands, respectively. In addition, an F160W-band image was obtained\cite{Levan2025} using the Wide Field Camera 3 (WFC3) on the Hubble Space Telescope (HST) through a Director's Discretionary Time program (ID: 17988, PI: Levan) on 15 July 2025. We measured magnitudes of $21.24 \pm 0.23$, $20.66 \pm 0.14$, and $21.21 \pm 0.07$ mag in the $H$ band, $K_s$ band, and F160W band, respectively. The photometry was obtained using an elliptical aperture with semi-major (semi-minor) axis of $1.75^{\prime \prime}$ ($0.65^{\prime \prime}$), and corrected for Galactic extinction. The $H$- and $K_s$-band magnitudes were calibrated against nearby 2MASS reference stars in the Vega system and then converted to the AB system by applying offsets of 1.39 ($H$) and 1.85 ($K_s$). Additionally, we inspected archival Pan-STARRS stacked images and found no detection of the host galaxy of EP250702a in the $z$ and $y$ bands, corresponding to upper limits of 21.8 and 21.1 mag in the $z$ and $y$ bands, respectively.

We performed a simple photometric redshift estimate for the host galaxy of EP250702a using this SED information with the Easy and Accurate Photometric Redshifts from Yale (EAZY\cite{EAZY}) code. We adopted the default v1.3 template set from EAZY, enabling the smoothing process by setting the "smoothed" parameter to "Y", while keeping all other parameters at their default values. The resulting maximum-likelihood redshift is $z = 0.604$, with a 1$\sigma$ confidence interval of 0.52--1.46. Using the Python package Code Investigating GALaxy Emission (CIGALE\cite{CIGALE}), we modeled the SED of the host galaxy at $z = 0.604$ and found the stellar mass of the galaxy is $10^{10.18\pm0.37}\,{M_{\odot}}$. The redshift estimation is consistent with the 
measurement of the redshift using the James Webb Space Telescope (JWST)\cite{gompertz2025jwst}.

\subsection{Afterglow modelling.\\}

We consider the afterglow emission of EP250702a arising from the interaction between a relativistic jet and the ambient medium. Since the X-ray emission is expected to be dominated by the early outburst phase powered by fallback accretion in the IMBH–WD system (see Theoretical modelling), we focus our modeling on the optical and radio data. Specifically, we employ the H- and K-band data collected by Ref.\cite{Levan2025}. Data in other optical bands are not used in the fitting, since only upper limits are available. We have checked these optical upper limits and find that all of them lie above our best-fit light curve, i.e., they do not provide additional constraints and are therefore fully consistent with the best-fit model. For the radio data, we make use of the multi-epoch observations from MeerKAT and GMRT at L-, S-, and C-band.

Following the approach of Ref.\cite{Huang..2000}, we calculate the jet dynamics and associated synchrotron radiation separately and fit the data with the MCMC algorithm (see also Ref.\cite{Xu..2023}). A redshift of $z=1.036$ is used in the fitting. The jet opening angle is fixed at $0.1$ rad, and is assumed to be viewed on-axis. The circumburst environment is modeled as a uniform interstellar medium. Since neither the optical nor the radio light curves exhibit a clear onset feature, it is hard to constrain the Lorentz factor though the afterglow modelling. Therefore, we fix the Lorentz factor at $56$ (see Constraints on the jet bulk Lorentz factor). A Milky Way–like extinction law is assumed for the host galaxy, characterized by $A_{V}$.

The best-fit light curves are shown in Fig.~\ref{fig:AG}a, where the model provides a good description of both the optical and radio data. 
The posterior distributions of the parameters are summarized in Fig.~\ref{fig:AG}b. 
The isotropic kinetic energy is constrained to be $4.1 \times 10^{53}$ erg. Our modeling yields a low circumburst density of $n \sim 0.001$ cm$^{-3}$, suggesting a relatively clean environment around the source. On the other hand, the high host galaxy extinction of $A_{V} \sim 4.6$ mag indicates a large dust column density along the line of sight, possibly due to a dense foreground region such as a molecular cloud.

\subsection{Theoretical modelling.\\}

The early flaring emission (ending with the rapid decay of the WXT flux) duration is $\sim$ 13 hours. This short duration disfavors a normal star as the disrupted object, whose time scale would be much longer (tens of days).
Compared to other sources, the X-ray flux in this event decays more slowly than typical GRBs but faster than the jetted TDE candidate Sw J1644+57. Combining these observational facts, the disrupted object is most likely a WD\cite{rosswog_tidal_2009,chen_tidal_2023}. Furthermore,  the strong jet implied by the GRB detection suggests the presence of an intense magnetic field. The disruption of a white dwarf naturally provides numerous seed magnetic fields within the accretion disk, supporting this interpretation.  

We employ the IMBH-WD TDE model to quantitatively interpret the observed light curve. Typically, a TDE light curve follows the mass fallback rate $\dot{M}_{\rm fb}$. We adopt the simulated fallback rate of a polytropic star with $5/3$ index from Ref.\cite{guillochon_hydrodynamical_2013} and rescale it for the IMBH-WD TDE scenario using Eqs.~(A1) and (A2) in their study. The X-ray emission arises from the accretion process, and its luminosity can be estimated as $L = \eta \dot{M}_{\rm acc} c^2$, where $\eta$ is the radiative efficiency of converting the accreted mass into X-ray emission. If the fallback mass is accreted by the IMBH without delay, the accretion rate $\dot{M}_{\rm acc}$ equals to the fallback rate $\dot{M}_{\rm fb}$. However, due to the uncertainties in disk formation and accretion physics in TDE, the accretion process is typically delayed. Following Ref.\cite{chen_tidal_2018}, we calculate the accretion rate by tracking the evolution of the disk mass, which is governed by the balance between mass supply from fallback and accretion (see also Refs.\cite{Kumar_Mass_2008,Lin_A_2017,Mockler_Weighing_2019}):
\begin{equation}
\frac{dM_{\rm disk}}{dt} = \dot{M}_{\rm fb} - \dot{M}_{\rm acc}.
\end{equation}
The accretion rate is assumed to follow: $\dot{M}_{\rm acc} \simeq M_{\rm disk}/\tau_{\rm acc}$, where $\tau_{\rm acc}$ is the viscous delay timescale. If $\tau_{\rm acc} \ll t_{\rm fb}$ (where $t_{\rm fb}$ is the fallback timescale), the accretion rate closely tracks the fallback rate ($\dot{M}_{\rm acc} \simeq \dot{M}_{\rm fb}$). Otherwise, the accretion process is prolonged by $\tau_{\rm acc}$.

Here we adopt the following mass-radius relationship of a WD \cite{Paczynski_Models_1983}:
\begin{equation} \label{eq:R_WD}
    R_* = 9 \times 10^8 \left[1-\left(\frac{M_*}{M_{\rm ch}}\right)^{4/3}\right]^{1/2} \left(\frac{M_*}{M_{\odot}}\right)^{-1/3},
\end{equation}
where $M_*$ is the WD mass and $M_{\rm ch} \simeq 1.44\ M_{\odot}$ is the Chandrasekhar mass.

We adopt a redshift of $z=1.036$ for this source. The observed isotropic X-ray luminosity is then given by $L_{\rm iso} \simeq F_{\rm X} 4\pi d^2$, where the corresponding luminosity distance $d\simeq 7093.76$ Mpc is derived assuming a flat {$\Lambda$}CDM cosmology.

Fig.~\ref{fig:IMBH_WD_modelling} compares our model light curve with observational data. The IMBH-WD TDE model successfully reproduces the observed emission, particularly the late-time power-law decay, as the thermal emission gradually dominates. Near the peak (where GRBs are detected), the sharp rise and decay of non-thermal emission suggest a relativistic jet, requiring beaming corrections. The observed $2-3$ order of magnitude discrepancy implies a jet Lorentz factor of $\Gamma \simeq 10-100$ during this phase, consistent with the constraint\cite{Oganesyan2025arXiv}. The model favors slowed accretion over rapid accretion, to match the data. The slowed accretion explains the delayed jet enhancement as well. In this assumed redshift, the radiative efficiency is $\eta \simeq 0.02$.

Notably, the accretion rate of a IMBH-WD TDE is extremely high, reaching up to 7 order of magnitude above the Eddington limit. The extreme Eddington ratio is different from the typical TDE, where accretion rate remain $\lesssim 100$ Eddington limit of accretion rate. Such extremely high accretion rate can lead to an extremely high luminosity in the thermal energy that exceeds the Eddington luminosity, which is suggested by some simulations of super-Eddington disks\cite{McKinney2015}. However, not all of the accretion power is converted into observable luminosity; our findings indicate that only a small fraction of the energy is radiated, with a radiative efficiency of approximately $\eta \simeq 0.02$.

An alternative model proposed to interpret the nature of this source involves repeating IMBH-WD TDEs. This model naturally accounts for the early extended, slow ascent and the subsequent rapid decline following the peak in the light curve, as illustrated in Fig.~\ref{fig:rTDE_WD_modelling}. Within this scenario, we assume that  a WD with an initial mass of $1.2\ M_{\odot}$ undergoes repeated tidal stripping by an IMBH of mass $10^4\ M_{\odot}$, starting in an eccentric orbit with impact factor $\beta \equiv R_{\rm t}/R_{\rm p} = 0.55$ and eccentricity $e_0=0.99$, corresponding to an initial orbital period of $P_0 \simeq 0.15$ days. Here $R_{\rm t} = R_*(M_{\rm h}/M_*)^{1/3}$ and $R_{\rm p}$ are the tidal radius and pericenter radius, respectively. We incorporate the change of orbital energy of the stellar remnant after each tidal stripping, adhering to the approach of Ref.\cite{chen_fate_2024}. Notably, the pericenter radius is unchanged, with only minor change in orbital energy. During each tidal stripping, we assume the mass-radius relation of the disrupted WD follows Eq.~(\ref{eq:R_WD}). The entire sequence comprises 4 partial TDEs followed by a final complete disruption. The model luminosity is given by $L=\eta \dot{M}_{\rm fb} c^2$ with efficiency of $\eta \simeq 0.04$.

When comparing these two models, the repeating IMBH-WD TDE model provides a natural interpretation for the delayed enhancement of the jet, since the final complete disruption involves a higher accreted mass and accretion rate. However, repeating TDEs are less probable than single TDEs, particularly as they require the initial orbit to only marginally graze the threshold for tidal stripping in order to generate multiple partial TDEs—a scenario with exceedingly low likelihood. Additionally, the absence of pronounced periodic variability in the light curve renders it inconsistent with the repeating TDE model.

\subsection{Event rate density.\\}

Offset TDEs serve as unique probes for wandering IMBHs, and their event rate density also provide valuable insights into the IMBH formation and evolution. At the  redshift of 1.036, the transient reaches an X-ray peak luminosity of approximately $\rm 3\times 10^{49}$ $\rm erg\,s^{-1}$ and maintains the peak time for a duration of one day. Since January 19, 2024, EP-WXT has been surveying the X-ray sky with an average cadence of 4.5 hours (3 orbits). Given the peak luminosity, the total maximum monitored volume-time is estimated at $3.9 \times 10^{12}$ $\rm Mpc^{3}\,yr$. As a result, the event rate density for transients similar to EP250702a is $2.6^{+5.9}_{-2.1} \times 10^{-13} $ $\rm Mpc^{-3}\,yr^{-1}$. This is more than two orders of magnetude lower compared with EP240222a\cite{Jin2025} and other known IMBH-TDE candidates\cite{Lin2018, Yao2025}, as well as the volumetric rate calculated for the off-nuclear IMBH-TDEs\cite{Chang2025ApJ}. 
This apparent discrepancy, however, aligns well with the expectation for a jetted origin scenario. For a jetted TDE with a small beaming factor $f_b \ll 1$, only a fraction of events are directed toward us and are detectable. The intrinsic event rate is therefore higher by a factor of $\sim f_b^{-1}$. Many similar events with jets not pointing toward us remain undetected. Finally, there should also be a population of more common, non-jetted IMBH-WD TDEs, though their fraction remains unknown. The X-ray emission of these non-jetted events is expected to be much weaker, making this luminous, jetted population distinct from those more isotropic and fainter events.

\subsection{Other possible scenarios.\\}

EP250702a's unique multi-wavelength characteristics, one-day-long X-ray peak as luminous as $10^{47-49}$\,erg\,s$^{-1}$ with strong recurrent flares with hard spectra extending to several tens of MeV gamma-rays, more than 5 orders of X-ray flux decay within 20 days, X-ray spectral softening at late times, pose a significant challenge to most conventional transient models. Several possible scenarios have been proposed to explain this extraordinary transient\cite{Beniamini2025, Neights2025}. Among these, the tidal disruption of a white dwarf by an IMBH (WD-IMBH TDE), as proposed herein and inferred by previous work on Sw J1644+57\cite{Krolik2011}, appears compelling. Rapid flaring activity in the high-energy band is reminiscent of blazars. However, they are long-lived sources of electromagnetic radiation, and none of them switches on and off on such short timescales. As already pointed out in the main text, the duration and spectral variability (strong softening) of the event is also inconsistent with classical known GRB (afterglows). An extreme kind of gravitational lensing event, where a compact source in the halo of the HST-detected galaxy (that would then be a foreground galaxy) lenses a background object is inconsistent with the monochromatic nature of gravitational lensing. Other mechanisms include micro-TDEs involving stellar-mass compact objects, a stellar mass black hole spiraling into a helium-star envelope, recently discussed in the context of ultra-long gamma-ray bursts. However, these models are difficult to explain the non-thermal, bright, and relatively stable X-ray emission observed approximately one day before the gamma-ray flare. Besides, given the significantly higher density of stellar mass black holes in the galactic plane, we would expect such events to arise preferentially from a galaxy's center, where the expected event rate should be much higher. Finally, exotic transients like white dwarf–neutron star collisions remain exceptionally rare and would likely produce prominent optical supernovae, which are not detected in this case.

\clearpage

\clearpage




\setcounter{table}{0}
\setcounter{figure}{0}

\renewcommand{\thefigure}{S\arabic{figure}}
\renewcommand{\thetable}{S\arabic{table}}
\captionsetup[figure]{name={\bf Fig.}}
\captionsetup[table]{name={\bf Table}}


\begin{figure}
\centering
\includegraphics[width=0.8\textwidth]{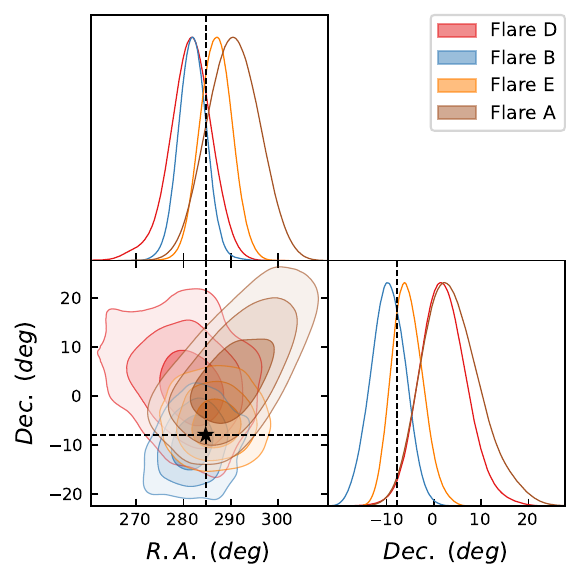}
\caption{Localization results of GRB 250702B derived from Fermi/GBM data using the BALROG method. The two-dimensional and marginalized one-dimensional localization probability distribution plots for Flares A (brown), D (red), B (blue), and E (orange) were generated using \texttt{GetDist}\cite{Lewis2025JCAP}. The filled contours in the two-dimensional plot represent the 1$\sigma$, and 2$\sigma$ and 3$\sigma$ statistical localization regions for each flare. The black star and dashed lines indicate the soft X-ray position of EP250702a, as determined by {\it Chandra}.}
\label{fig:gbm_balrog}
\end{figure}

\clearpage

\begin{figure}
\centering
\begin{tabular}{cc}
\begin{overpic}[width=0.45\textwidth]{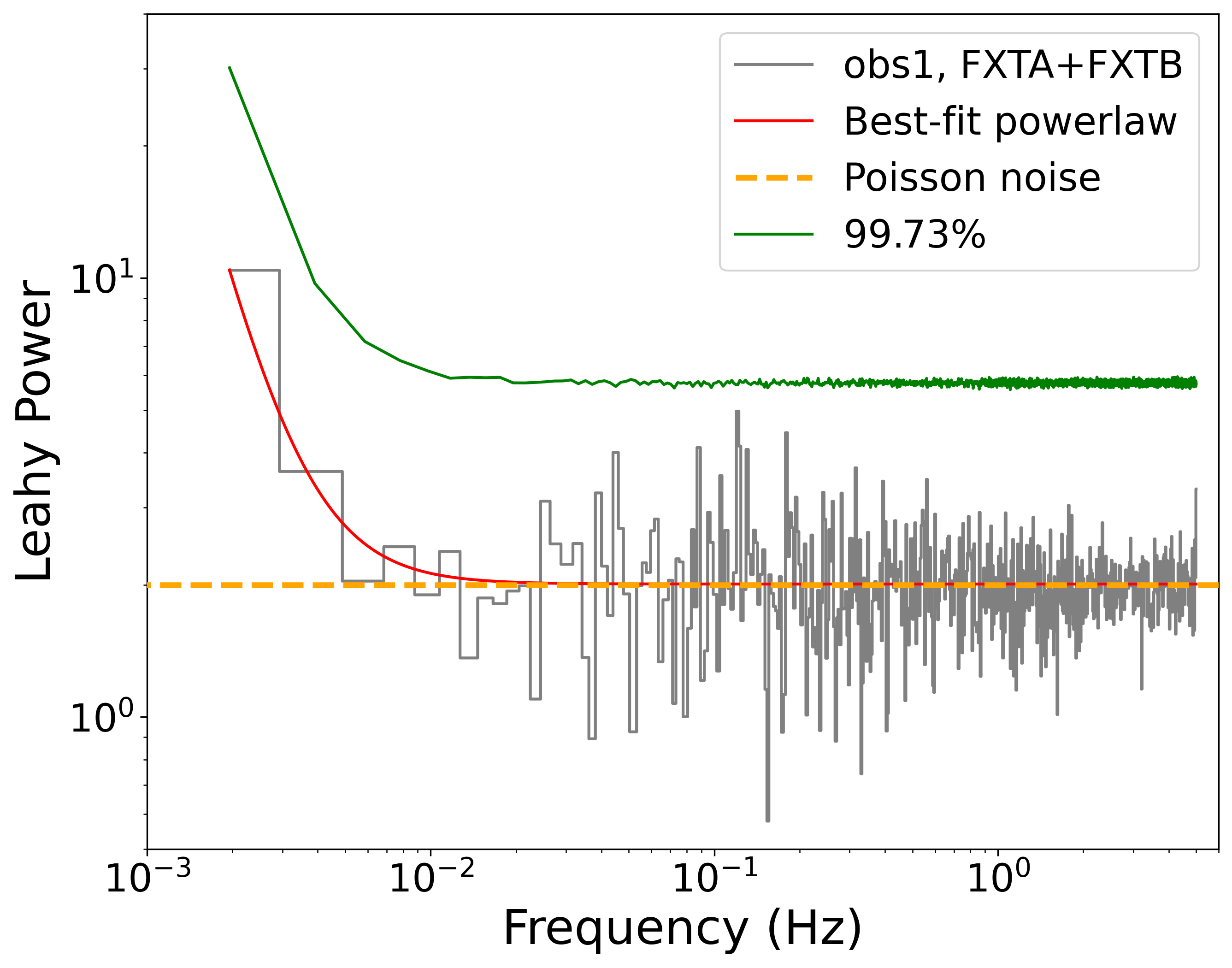}\put(0, 80){\bf (a)}\end{overpic} &
\begin{overpic}[width=0.45\textwidth]{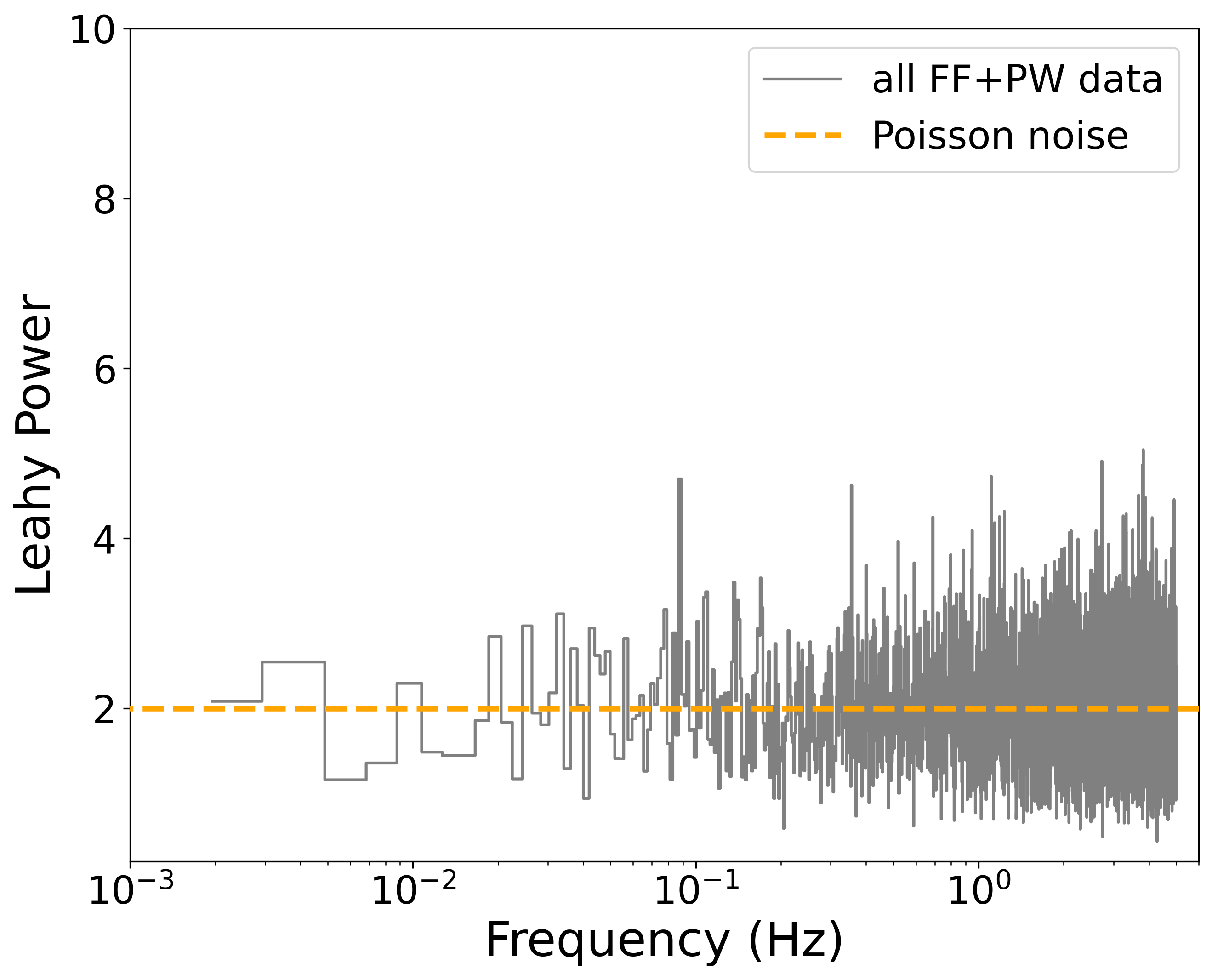}\put(0, 80){\bf (b)}\end{overpic} \\
\end{tabular}
\caption{EP250702a’s X-ray power spectral density (PSD). (a), The data were taken from the first ToO observation of EP-FXT carried out on 2025-07-03, the flux of which is the highest among all follow-up observations. FXTA and FXTB data are combined for achieving a higher signal-to-noise ratio. The data is denoted in gray histogram, the best-fit power-law model is denoted in red solid line, and the Poisson noise is denoted in orange dashed line. The green solid line is the PSD at $3\sigma$ confidence level obtained from 100000 simulated light curves from the best-fitting power-law PSD. No potential QPO signal is observed. (b), All the data taken from imaging mode (FF+PW) from 2025-07-20T14:50:24 to 2025-07-24T10:39:27 were combined for a stacked PSD. The PSD is dominated by white noise at all measured frequencies.}
\label{fig:fxt_obs1_psd}
\end{figure}

\clearpage

\begin{figure}
\centering
\begin{tabular}{cc}
\begin{overpic}[width=0.45\textwidth]{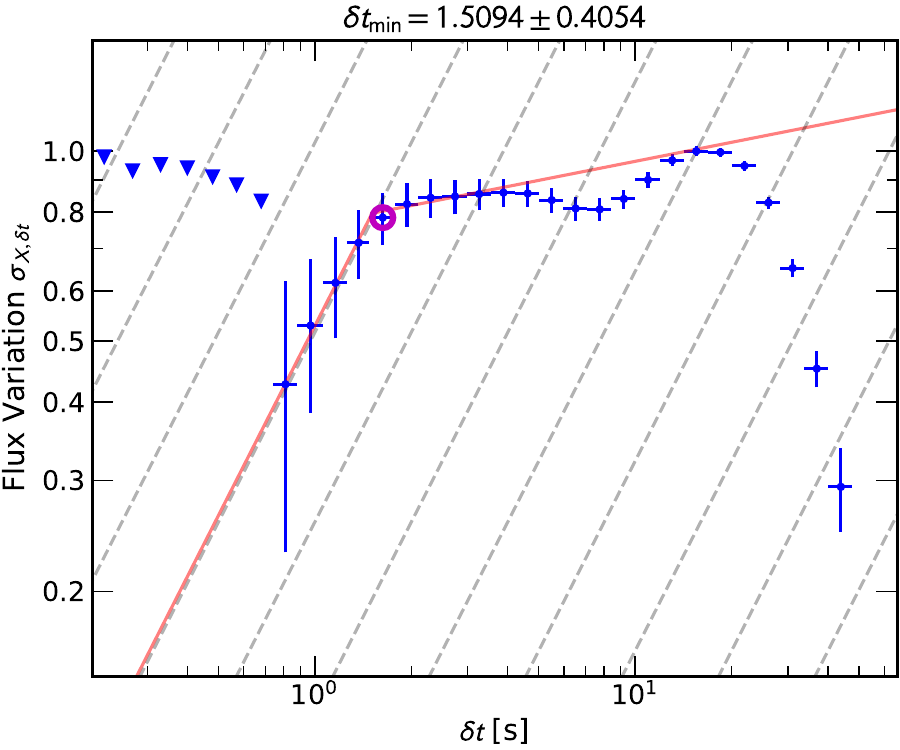}\put(0, 80){\bf (a)}\end{overpic} &
\begin{overpic}[width=0.45\textwidth]{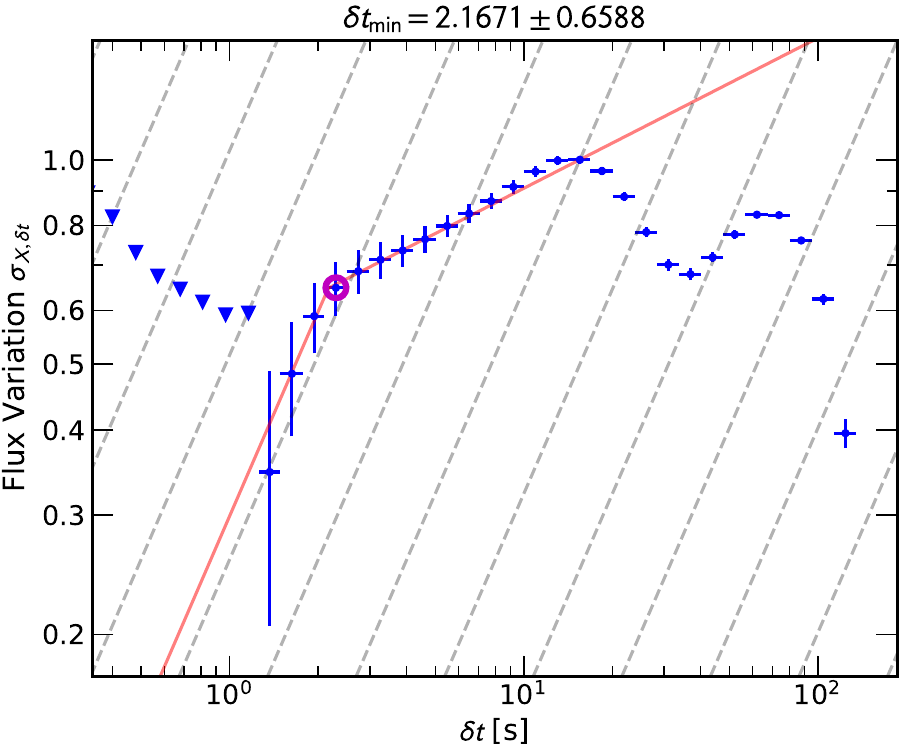}\put(0, 80){\bf (b)}\end{overpic} \\
\end{tabular}
\caption{The minimum variability timescale of GRB 250702B. Flux variation as a function of variability timescale derived from Fermi/GBM data within the time intervals $T_{\rm 0, D} + [-95, 10]$ (a) and $T_{\rm 0, E} + [-5, 180]$ (b). The red solid line indicates the transition from the temporally smooth region ($\sigma_{X, \delta t} \propto \delta t$; gray dashed lines) to a flatter regime, while the magenta circle marks the break point corresponding to the minimum variability timescale ($\delta t_{\rm min}$). Triangles denote 3$\sigma$ upper limits.}
\label{fig:gbm_mvt}
\end{figure}

\clearpage

\begin{figure}
\centering
\begin{tabular}{cc}
\begin{overpic}[width=0.45\textwidth]{figures/sup_fig4a.pdf}\put(-3, 95){\bf (a)}\end{overpic} &
\begin{overpic}[width=0.45\textwidth]{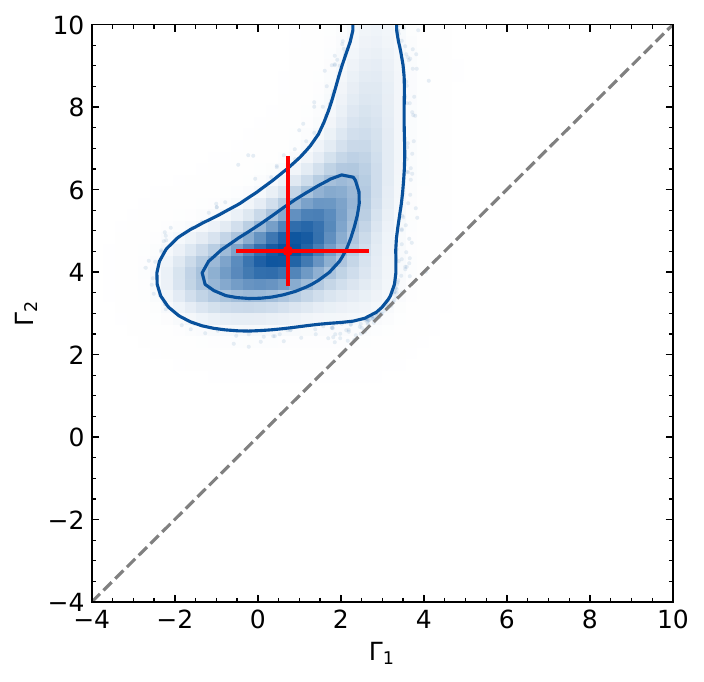}\put(-3, 95){\bf (b)}\end{overpic} \\
\end{tabular}
\caption{(a), Joint spectral fitting of simultaneous EP-FXT and {\it Chandra} observations. The best-fit absorbed power-law model components (dashed and dotted lines), as well as their combined model \textit{const*tbabs*(powerlaw+powerlaw)} (solid line), are shown in the top panel. The fitting residuals are shown in the lower panel. The comparative results for a single absorbed power-law model shown in the middle panel indicate systematic deviations above 2\,keV. All error bars represent 1$\sigma$ uncertainties. (b), Posterior probability distributions of the photon indices for the two power-law models, derived via Bayesian inference. The contours denote the 1$\sigma$ and the 2$\sigma$ confidence levels. The red error bars represent the 1$\sigma$ uncertainties, and the red dot indicates the best-fit values.}
\label{fig:joint_spec}
\end{figure}

\clearpage

\begin{figure}
\centering
\includegraphics[width=0.8\textwidth]{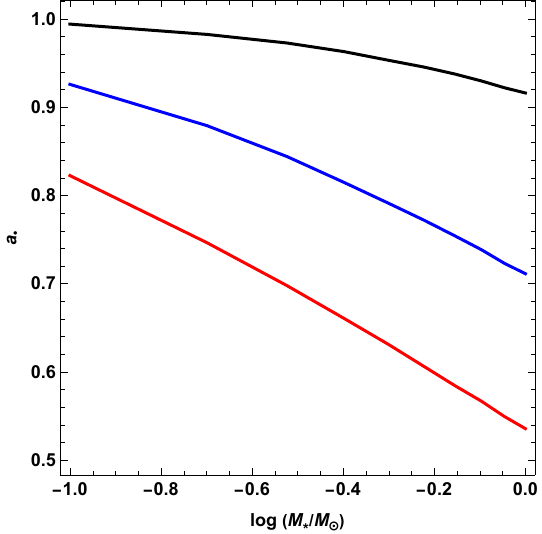}
\caption{Parameter space of $M_* -  a_{\bullet}$ for EP250702a. The red line indicate the range for BH spin with $f_b = 1.6\times 10^{-4}$. The blue line corresponds to the value $f_b=5.9\times 10^{-4}$. The black line corresponds to the value $f_b=5\times 10^{-3}$. It is shown that the  probable values are $a_{\bullet}  = 0.54, 0.82$ (red line), $a_{\bullet}  = 0.71, 0.93$ (blue line), and $a_{\bullet}  = 0.92, 0.99$ (black line) for $M_* = 1.0, 0.1 M_\odot$ , respectively. $\eta=0.5$ is adopted in the calculations. }
\label{fig:BH_spin}
\end{figure}

\clearpage

\begin{figure}
\centering
\begin{tabular}{cc}
\begin{overpic}[width=0.52\textwidth]{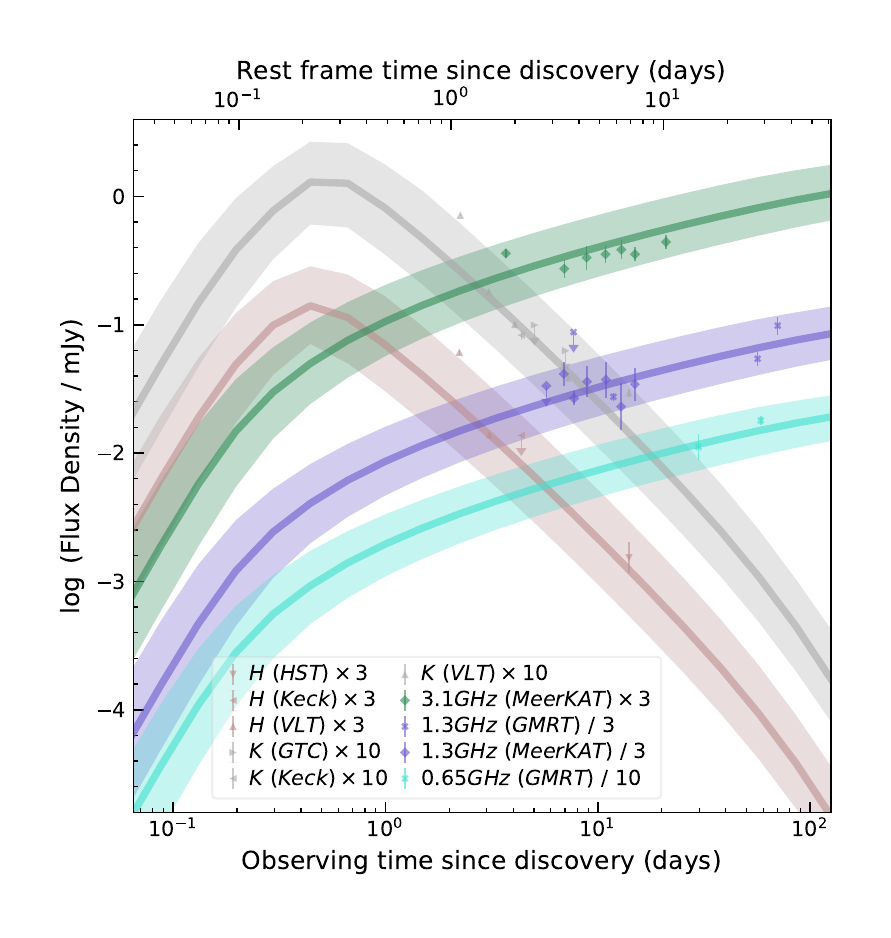}\put(0, 90){\bf (a)}\end{overpic} &
\begin{overpic}[width=0.50\textwidth]{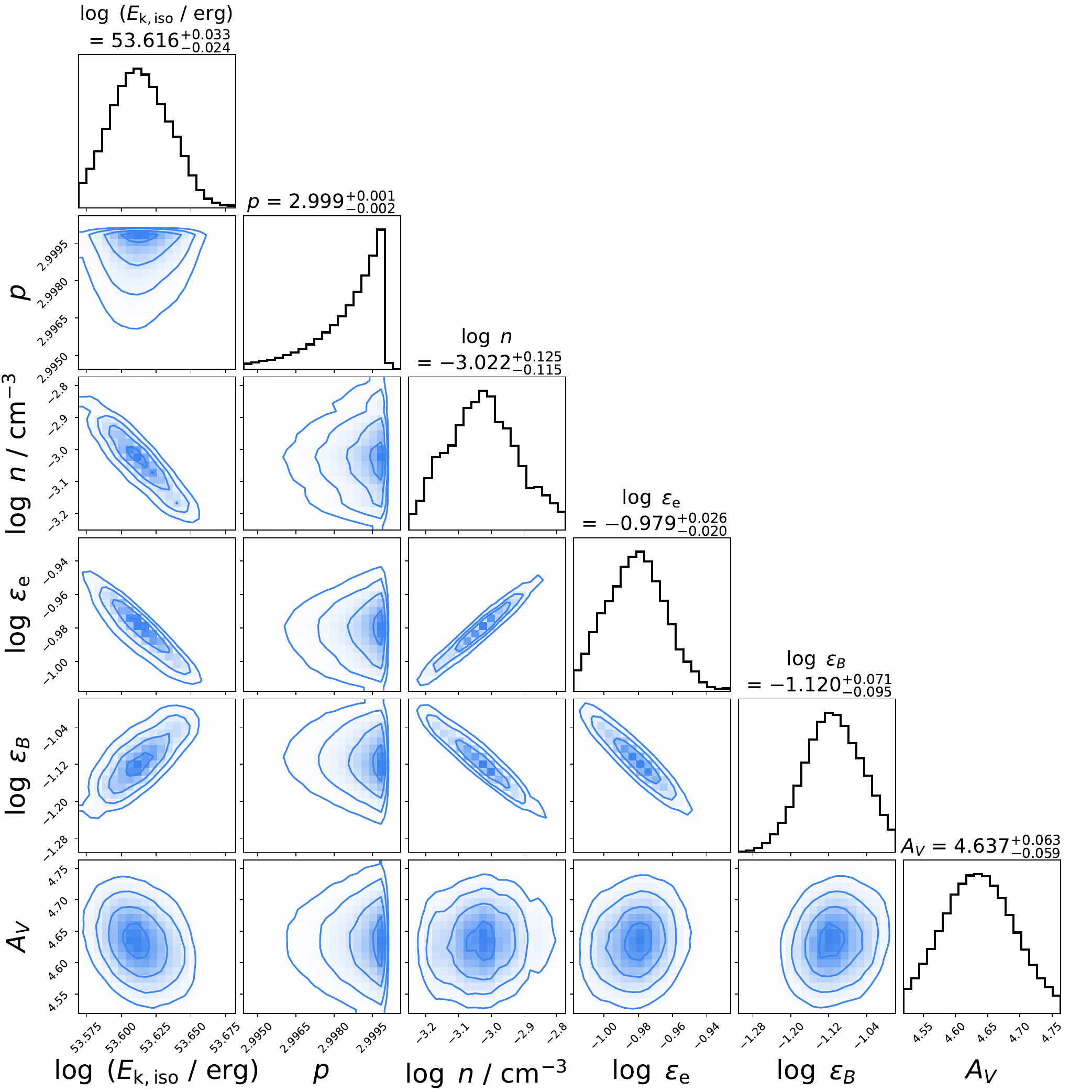}\put(0, 99){\bf (b)}\end{overpic} \\
\end{tabular}
\caption{Afterglow modelling of EP250702a. (a), Multi-wavelength afterglow data of EP250702a compared with the corresponding model light curves (solid lines). Shaded regions exhibit the $1\sigma$ confidence intervals of the model light curves derived from posterior distributions of model parameters. (b), Posterior constraints on the afterglow model parameters of EP250702a. The contours mark the $1\sigma$, $2\sigma$, and $3\sigma$ confidence regions, and the best-fit values with their $1\sigma$ uncertainties are listed above the corresponding posterior distributions.}
\label{fig:AG}
\end{figure}

\clearpage

\begin{figure}
\centering
\begin{tabular}{cc}
\includegraphics[width=1\textwidth]{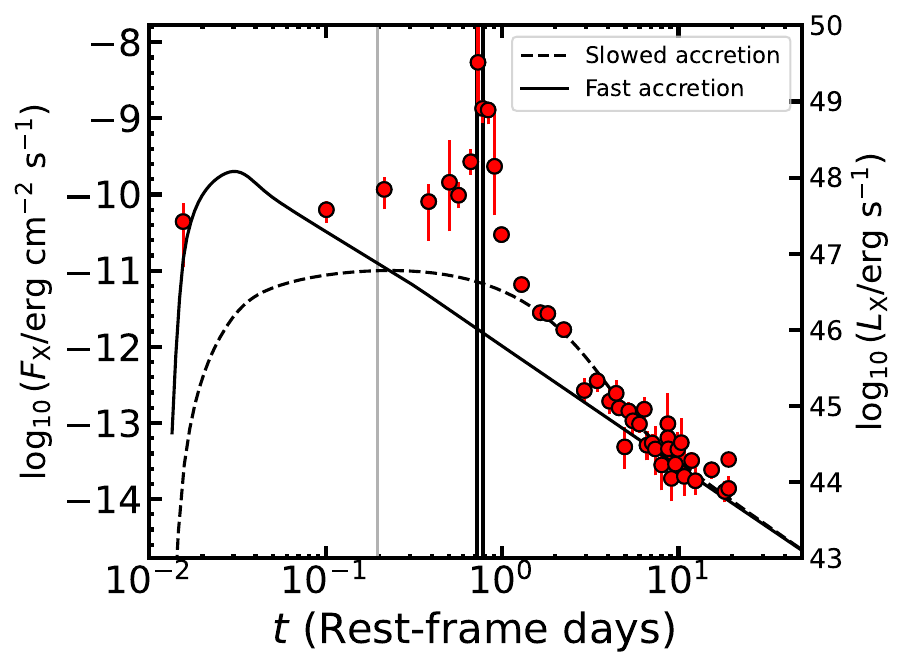}
\end{tabular}
\caption{IMBH-WD TDE modelling and the X-ray isotropic luminosity in 0.3--10 KeV band. The solid lines indicate the luminosity corresponding to an accretion rate equal to the mass fallback rate, whereas the dashed lines represent scenarios where the accretion rate is delayed by a timescale $\tau_{\rm acc} \simeq 0.8$ day. The vertical black and gray lines denote the trigger times of the Fermi flares and untriggered weak Fermi flare, respectively. The mass of disrupted WD is $0.4\ M_{\odot}$. The model luminosity is given by $L=\eta \dot{M}_{\rm acc} c^2$, where the efficiency is $\eta \simeq 0.02z^2$. The right Y-axis denotes the isotropic luminosity of the observational data with the assumed redshift $z=1.036$.}
\label{fig:IMBH_WD_modelling}
\end{figure}

\clearpage

\begin{figure}
\centering
\begin{tabular}{cc}
\includegraphics[width=1\textwidth]{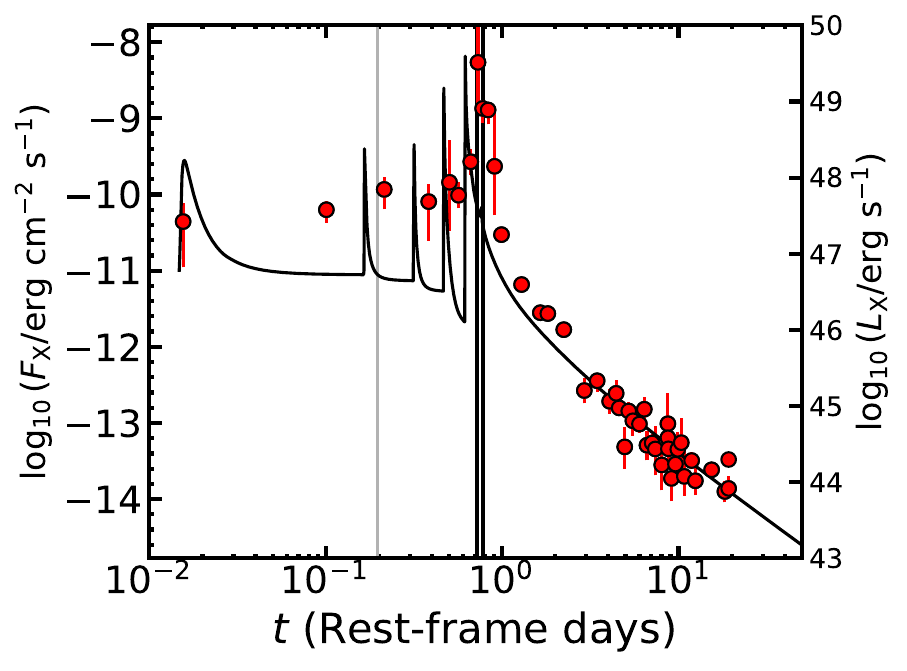}
\end{tabular}
\caption{Modelling of repeating IMBH-WD TDE and the X-ray isotropic luminosity in 0.3--10 KeV band. The vertical black and gray lines denote the trigger times of the Fermi flares and untriggered weak Fermi flare, respectively. The disrupted WD has an initial mass of $1.2\ M_{\odot}$ and starts in an eccentric orbit with $\beta = 0.55$ and $e_0=0.99$, corresponding to an initial orbital period of $P_0 \simeq 0.15$ days. The event comprises 4 partial TDEs followed by a final complete disruption. The model luminosity is given by $L=\eta \dot{M}_{\rm fb} c^2$, where the efficiency is $\eta \simeq 0.04z^2$. The right Y-axis denotes the isotropic luminosity of the observational data with the assumed redshift $z=1.036$.}
\label{fig:rTDE_WD_modelling}
\end{figure}

\clearpage

{\renewcommand{\arraystretch}{0.85}
\begin{table*}
\footnotesize
\centering
\caption{\noindent\textbf{X-ray observational log and spectral fitting of EP250702a.} For all spectral modeling, the Galactic absorption was fixed at $3.3\times10^{21}$ cm$^{-2}$. The $N_{\rm H}$ values given in the table represent the intrinsic absorption column densities derived from the fits. All errors represent the 1$\sigma$ uncertainties.}
\label{tab:xray_label}
\begin{tabular}{cccccc}
\hline
Observational time & ObsID & Exposure & $N_{\rm H}$ & $\Gamma$ & log${F_{\rm abs, 0.3-10~keV}}$ \\
(UTC) & & (s) & ($10^{22}$~${\rm cm}^{-2}$) & & (${\rm erg~cm^{-2}~s^{-1}}$) \\
\hline
& EP-WXT & & & & \\
2025-07-02T02:53:45 & 01709180547 & 3397 & -& $0.3^{+0.7}_{-0.7}$& $-9.8_{-0.6}^{+0.6}$  \\
2025-07-02T05:58:17 & 06800000718 & 5377 & $3.0^{+3.7}_{-2.5}$ & $1.6^{+0.8}_{-0.6}$ & $-10.0_{-0.2}^{+0.2}$ \\
2025-07-02T10:45:53 & 06800000712 & 6126 & $2.8^{+2.6}_{-1.9}$ & $1.1^{+0.5}_{-0.5}$ & $-9.6_{-0.2}^{+0.2}$ \\
2025-07-02T13:57:44 & 11900293634 & 152 & - & $-0.6^{+1.1}_{-1.4}$& $-8.3_{-0.6}^{+0.7}$ \\
2025-07-02T16:11:21 & 11916649980 & 827 & $0.2^{+3.9}_{-0.2}$ &$1.1^{+0.5}_{-0.5}$ & $-8.9_{-0.2}^{+0.1}$ \\ 
2025-07-02T18:59:25 & 06800000715 & 4267 & 0.2$_{-0.2}^{+2.0}$ &$0.1^{+0.5}_{-0.2}$ & $-8.9_{-0.2}^{+0.1}$ \\
2025-07-02T22:33:54 & 11900293762 & 816 & - & $1.3^{+0.7}_{-0.6}$& $-9.6_{-0.6}^{+0.7}$ \\
2025-07-03T02:46:32 & 06800000725 & 3897 & - & $0.8_{-0.5}^{+0.5}$& $-9.9_{+0.2}^{-0.3}$ \\ 
\hline
& EP-FXT & & & & \\
2025-07-03T02:45:21 & 06800000725 & 3876 & $2.6_{-0.2}^{+0.2}$&$1.56_{-0.05}^{+0.05}$& $-10.53_{-0.02}^{+0.02}$ \\
2025-07-03T17:16:25 & 06800000727 & 4765 & $1.8_{-0.7}^{+0.7}$ &$1.5_{-0.1}^{+0.1}$& $-11.18_{-0.04}^{+0.04}$ \\
2025-07-04T10:44:38 & 06800000728 & 3015 & $1.2_{-1.1}^{+1.3}$&$1.4_{-0.3}^{+0.3}$& $-11.55_{-0.09}^{+0.09}$ \\
2025-07-04T18:57:58 & 06800000729 & 4133 & $3.4_{-1.5}^{+1.7}$ &$1.6_{-0.3}^{+0.3}$& $-11.56_{-0.09}^{+0.09}$ \\
2025-07-05T15:31:56 & 06800000733 & 7760 & $5.1_{-1.6}^{+1.9}$ &$1.9_{-0.3}^{+0.3}$& $-11.78_{-0.07}^{+0.07}$ \\
2025-07-07T01:07:06 & 06800000738 & 3902 & $3.9_{-2.8}^{+4.1}$ &$2.4_{-0.7}^{+0.9}$& $-12.6_{-0.2}^{+0.2}$ \\
2025-07-08T03:12:21 & 11900301312 & 5807 & $0.1_{-0.1}^{+2.1}$ &$1.8_{-0.3}^{+0.5}$& $-12.4_{-0.1}^{+0.1}$ \\
2025-07-09T09:05:43 & 11900303104 & 5126 & $4.5_{-2.7}^{+3.6}$ &$2.6_{-0.7}^{+0.9}$& $-12.7_{-0.2}^{+0.2}$ \\
2025-07-10T02:41:19 & 06800000743 & 5385 & $1.1_{-1.1}^{+2.1}$ &$2.0_{-0.5}^{+0.7}$& $-12.6_{-0.2}^{+0.2}$\\
2025-07-10T11:20:05 & 11900304256 & 5939 & $-$ &$2.4_{-0.6}^{+0.6}$& $-12.8_{-0.2}^{+0.2}$\\
2025-07-11T04:16:45 & 06800000746 & 2901 & $-$ &$2.7_{-2.2}^{+1.9}$& $-13.2_{-0.5}^{+0.9}$ \\
2025-07-11T17:10:58 & 08500000365 & 7299 & $-$ &$2.2_{-0.5}^{+0.5}$& $-12.8_{-0.2}^{+0.2}$ \\
2025-07-12T07:28:14 & 06800000749 & 5987 & $-$ &$1.9_{-0.8}^{+0.8}$& $-12.9_{-0.4}^{+0.3}$ \\
2025-07-13T07:27:52 & 06800000751 & 9151 & $-$ &$2.6_{-0.6}^{+0.6}$& $-13.0_{-0.2}^{+0.2}$ \\
2025-07-14T02:41:25 & 06800000752 & 6057 & $-$ &$1.8_{-0.7}^{+0.8}$& $-12.8_{-0.3}^{+0.3}$ \\
2025-07-14T13:53:16 & 06800000756 & 6071 & $-$ &$2.3_{-0.9}^{+1.0}$& $-13.2_{-0.4}^{+0.3}$ \\
2025-07-15T12:17:03 & 06800000757 & 7353 & $-$ &$2.9_{-1.3}^{+1.1}$& $-13.2_{-0.3}^{+0.4}$ \\
2025-07-16T02:58:36 & 11900311296 & 6204 & $-$&$3.0_{-1.8}^{+1.8}$& $-13.2_{-0.4}^{+0.5}$ \\
2025-07-17T10:40:51 & 06800000765 & 6179 & $-$ &$3.6_{-1.8}^{+2.4}$& $-13.3_{-0.3}^{+0.3}$\\
2025-07-18T21:59:59 & 08500000366 & 11304 &$-$&$3.8_{-0.9}^{+0.9}$& $-13.3_{-0.2}^{+0.1}$\\
2025-07-19T15:29:04 & 06800000766 & 9330 & $-$ &$2.2_{-2.4}^{+1.8}$& $-13.3_{-0.5}^{+1.3}$\\
2025-07-20T15:29:16 & 06800000770 & 9262 & $-$ &$2.7_{-1.5}^{+1.5}$& $-13.6_{-0.9}^{+0.5}$\\
2025-07-21T05:53:23 & 06800000772 & 8697 & $-$ &$2.6_{-1.3}^{+1.3}$& $-13.5_{-0.4}^{+0.4}$\\
2025-07-22T04:17:34 & 06800000774 & 8770 & $-$ &$3.1_{-1.3}^{+1.3}$ & $-13.5_{-0.4}^{+0.3}$\\
2025-07-23T01:05:41 & 11900318720 & 8542 & $-$ &$3.6_{-2.0}^{+2.3}$& $-13.8_{-0.5}^{+0.5}$\\
2025-07-25T05:52:34 & 08500000371 & 15723& $-$ &$2.3_{-1.4}^{+1.3}$& $-13.6_{-0.5}^{+0.4}$\\
2025-07-26T10:40:22 & 08500000372 & 23652& $-$ &$2.6_{-1.3}^{+1.5}$& $-13.7_{-0.3}^{+0.2}$\\
2025-08-01T10:36:40 & 08500000374 & 23659& $-$ &$3.2_{-1.2}^{+1.2}$& $-13.7_{-0.3}^{+0.3}$\\
2025-08-07T08:56:39 & 08500000377 & 23291& $-$ &$2.9_{-0.8}^{+0.8}$& $-14.0_{-0.3}^{+0.3}$\\
2025-08-14T07:15:11 & 08500000379 & 22653& $-$ &$3.7_{-1.1}^{+1.2}$& $-13.7_{-0.2}^{+0.3}$\\
2025-08-18T07:12:39 & 08500000380 & 22214& $-$ &$4.5_{-1.4}^{+1.5}$& $-13.9_{-0.3}^{+0.3}$\\
\hline
& Chandra/ACIS & & & & \\
2025-07-18T19:24:33 & 31003 & 14900 &$-$ &$2.6_{-0.4}^{+0.4}$& $-13.21_{-0.07}^{+0.07}$ \\
\hline
\end{tabular}
\end{table*}}

\clearpage

\begin{table*}
\centering
\caption{\noindent\textbf{Log of GMRT observations of EP250702a used in this work.}}
\label{tab:radio_obs}
\begin{tabular}{ccccc}
\hline
Observational time & Telescope & Frequency & Peak flux & MAP\_RMS \\
(UTC) & & (GHz) & ($\mu$Jy) & ($\mu$Jy/beam) \\
\hline
2025-07-08T18:00:00 & GMRT & 1.264 & <262.9 & 87.64  \\
2025-07-12T21:30:00 & GMRT & 1.264 &  82.3  & 05.94 \\
2025-07-30T19:30:00 & GMRT & 0.647 &  111.8 & 26.11 \\
2025-08-26T19:30:00 & GMRT & 1.264 &  163.0 & 20.69 \\
2025-08-28T18:30:00 & GMRT & 0.647 &  179.4 & 14.31 \\
2025-09-09T14:00:00 & GMRT & 1.264 &  295.0 & 47.64 \\
\hline
\end{tabular}
\end{table*}

\clearpage

\begin{table*}
\centering
\caption{\noindent\textbf{Best-fit parameters from the joint spectral fitting of the simultaneous EP-FXT and Chandra Observations.} All uncertainties are reported at the $1\sigma$ confidence level. The spectral fitting indicates that no intrinsic absorption beyond the Galactic component is required.}
\label{tab:joint_fit_xray}
\begin{tabular}{cccccc}
\hline
Model & $\Gamma_1$ & $\Gamma_2$ & $T_{\rm bb}$ (eV) & const & CSTAT/d.o.f. \\
\hline
\textit{const*tbabs*powerlaw} & 2.9$^{+0.4}_{-0.4}$ & - & -& 0.87 & 82.42/115  \\
\textit{const*tbabs*(powerlaw+powerlaw)} & 0.7$^{+1.2}_{-2.3}$ & 4.75$^{+1.5}_{-1.0}$ &-& 0.76& 74.89/113 \\
\textit{const*tbabs*(bbody+powerlaw)} & 1.4$^{+0.8}_{-0.8}$ & - & 169$^{+37}_{-37}$  & 0.78 & 74.89/113 \\
\hline
\end{tabular}
\end{table*}

\clearpage

\begin{table*}
\centering
\caption{\noindent\textbf{Localization results of GRB 250702B derived from Fermi/GBM data using the BALROG method.} All errors represent the 2$\sigma$ uncertainties.}
\label{tab:gbm_balrog}
\begin{tabular}{cccc}
\hline
Flare & Time interval & Right ascension & Declination \\
& (s) & (degree) & (degree) \\
\hline
Flare A & $T_{\rm 0, A} + [-5, 5]$ & $288_{-7}^{+14}$ & $2_{-9}^{+15}$ \\
Flare D & $T_{\rm 0, D} + [-53, -43]$ & $282_{-11}^{+8}$ & $2_{-9}^{+11}$ \\
Flare B & $T_{\rm 0, B} + [-6, 4]$ & $282_{-5}^{+6}$ & $-8_{-9}^{+5}$ \\
Flare E & $T_{\rm 0, E} + [127, 137]$ & $285_{-5}^{+9}$ & $-6_{-5}^{+8}$ \\
\hline
\end{tabular}
\end{table*}

\clearpage

\begin{table*}
\centering
\setlength\tabcolsep{4pt}
\scriptsize
\caption{\noindent\textbf{The temporal and spectral characteristics during the flaring episodes of EP250702a.} The models marked with a \checkmark symbol represent the best-fitting models, as indicated by the smallest BIC value. The unabsorbed flux is derived from the best fitting model within the energy range of 1--10000 keV. All errors represent the 1$\sigma$ uncertainties.}
\label{tab:x_gamma_prop}
\begin{tabular}{ccclccccc}
\hline
Time interval & $\delta t_{\rm min}$ & Unabsorbed flux & Model & $\alpha$ & $\beta$ & log$E_{\rm p}$ & STAT/d.o.f. & BIC \\
(s) & (s) & (${\rm erg~cm^{-2}~s^{-1}}$) & & & & (keV) & & \\
\hline
\multirow{4}{*}{\makecell{SED1 \\ $T_{\rm 0, A} + [-10, 20]$}} & \multirow{4}{*}{$3.7\pm0.6$} & \multirow{4}{*}{$6.24_{-1.02}^{+1.23}\times10^{-8}$} & \textit{pl} & $-1.78_{-0.06}^{+0.06}$ & -- & -- & $436.96/354$ & $448.71$ \\
& & & \textit{cpl}\checkmark & $-1.21_{-0.25}^{+0.27}$ & -- & $1.91_{-0.10}^{+0.16}$ & $428.30/353$ & $445.93$ \\
& & & \textit{band} & $-1.22_{-0.27}^{+0.43}$ & unconstrained & $1.90_{-0.12}^{+0.20}$ & $427.67/352$ & $451.17$ \\
& & & \textit{sbpl} & $-1.50_{-0.19}^{+0.30}$ & unconstrained & $1.92_{-0.17}^{+0.34}$ & $429.08/352$ & $452.15$ \\
\hline
\multirow{4}{*}{\makecell{SED2 \\ $T_{\rm 0, D} + [-95, 10]$}} & \multirow{4}{*}{$1.5\pm0.4$} & \multirow{4}{*}{$1.64_{-0.33}^{+0.34}\times10^{-7}$} & \textit{tbabs*ztbabs*pl} & $-1.33_{-0.02}^{+0.02}$ & -- & -- & $321.38/238$ & $332.35$ \\
& & & \textit{tbabs*ztbabs*cpl}\checkmark & $-0.87_{-0.08}^{+0.08}$ & -- & $2.81_{-0.12}^{+0.12}$ & $241.40/237$ & $257.84$ \\
& & & \textit{tbabs*ztbabs*band} & $-0.86_{-0.08}^{+0.09}$ & $-2.78_{-0.79}^{+0.56}$ & $2.79_{-0.13}^{+0.14}$ & $240.36/236$ & $262.28$ \\
& & & \textit{tbabs*ztbabs*sbpl} & $-0.92_{-0.07}^{+0.08}$ & $-2.42_{-0.59}^{+0.27}$ & $2.66_{-0.14}^{+0.19}$ & $239.49/236$ & $261.41$ \\
\hline
\multirow{4}{*}{\makecell{SED3 \\ $T_{\rm 0, B} + [-40, 40]$}} & \multirow{4}{*}{$<4.2$} & \multirow{4}{*}{$3.66_{-0.53}^{+0.42}\times10^{-7}$} & \textit{pl} & $-1.38_{-0.02}^{+0.02}$ & -- & -- & $414.49/234$ & $425.42$ \\
& & & \textit{cpl} & $-0.72_{-0.16}^{+0.17}$ & -- & $2.78_{-0.11}^{+0.18}$ & $368.49/233$ & $384.88$ \\
& & & \textit{band} & $-0.59_{-0.19}^{+0.26}$ & $-2.05_{-0.12}^{+0.04}$ & $2.68_{-0.13}^{+0.17}$ & $353.46/232$ & $375.12$ \\
& & & \textit{sbpl}\checkmark & $-0.74_{-0.14}^{+0.16}$ & $-2.06_{-0.09}^{+0.04}$ & $2.87_{-0.19}^{+0.25}$ & $352.32/232$ & $373.73$ \\
\hline
\multirow{4}{*}{\makecell{SED4 \\ $T_{\rm 0, B} + [100, 190]$}} & \multirow{4}{*}{$11\pm4$} & \multirow{4}{*}{$1.12_{-0.30}^{+0.40}\times10^{-7}$} & \textit{tbabs*ztbabs*pl} & $-1.30_{-0.03}^{+0.03}$ & -- & -- & $434.90/238$ & $445.86$ \\
& & & \textit{tbabs*ztbabs*cpl}\checkmark & $-0.86_{-0.11}^{+0.14}$ & -- & $2.80_{-0.17}^{+0.22}$ & $408.68/237$ & $425.12$ \\
& & & \textit{tbabs*ztbabs*band} & $-0.86_{-0.12}^{+0.15}$ & $-2.65_{-0.89}^{+0.51}$ & $2.80_{-0.17}^{+0.21}$ & $406.51/236$ & $428.43$ \\
& & & \textit{tbabs*ztbabs*sbpl} & $-0.92_{-0.11}^{+0.12}$ & $-2.56_{-0.77}^{+0.38}$ & $2.67_{-0.16}^{+0.20}$ & $406.14/236$ & $428.06$ \\
\hline
\multirow{4}{*}{\makecell{SED5 \\ $T_{\rm 0, E} + [-330, -170]$}} & \multirow{4}{*}{$3.6\pm1.7$} & \multirow{4}{*}{$3.54_{-0.65}^{+0.50}\times10^{-7}$} & \textit{tbabs*ztbabs*pl} & $-1.32_{-0.02}^{+0.02}$ & -- & -- & $320.35/238$ & $331.31$ \\
& & & \textit{tbabs*ztbabs*cpl}\checkmark & $-1.03_{-0.04}^{+0.04}$ & -- & $3.32_{-0.11}^{+0.10}$ & $212.81/237$ & $229.25$ \\
& & & \textit{tbabs*ztbabs*band} & $-1.03_{-0.04}^{+0.05}$ & $-3.93_{-1.26}^{+1.19}$ & $3.31_{-0.12}^{+0.11}$ & $212.73/236$ & $234.65$ \\
& & & \textit{tbabs*ztbabs*sbpl} & $-1.09_{-0.03}^{+0.05}$ & $-2.91_{-1.28}^{+0.69}$ & $3.27_{-0.14}^{+0.11}$ & $216.00/236$ & $237.92$ \\
\hline
\multirow{4}{*}{\makecell{SED6 \\ $T_{\rm 0, E} + [-5, 180]$}} & \multirow{4}{*}{$2.2\pm0.6$} & \multirow{4}{*}{$2.94_{-0.31}^{+0.16}\times10^{-7}$} & \textit{tbabs*ztbabs*pl} & $-1.34_{-0.01}^{+0.01}$ & -- & -- & $551.62/239$ & $562.59$ \\
& & & \textit{tbabs*ztbabs*cpl} & $-0.86_{-0.05}^{+0.05}$ & -- & $2.80_{-0.07}^{+0.07}$ & $329.83/238$ & $346.29$ \\
& & & \textit{tbabs*ztbabs*band} & $-0.83_{-0.06}^{+0.06}$ & $-2.10_{-0.18}^{+0.07}$ & $2.73_{-0.09}^{+0.08}$ & $318.16/237$ & $340.10$ \\
& & & \textit{tbabs*ztbabs*sbpl}\checkmark & $-0.85_{-0.06}^{+0.06}$ & $-2.05_{-0.06}^{+0.03}$ & $2.89_{-0.13}^{+0.17}$ & $315.65/237$ & $337.59$ \\
\hline
\end{tabular}
\end{table*}